\documentclass[fleqn,usenatbib]{mnras}

\usepackage{newtxtext,newtxmath}

\usepackage[T1]{fontenc}

\DeclareRobustCommand{\VAN}[3]{#2}
\let\VANthebibliography\thebibliography
\def\thebibliography{\DeclareRobustCommand{\VAN}[3]{##3}\VANthebibliography}

\usepackage{graphicx}	
\usepackage{amsmath}	
\usepackage{gensymb}
\usepackage{threeparttable}

\newcommand{\msun}{\mathrm{M}\textsubscript{\(\odot\)}}
\newcommand{\rsun}{\mathrm{R}\textsubscript{\(\odot\)}}

\defcitealias{Tailo22_M4}{T22}
\defcitealias{Tailo20_massloss_difference_multipops}{T20}
\defcitealias{Origlia14_masslossGCs}{O14}
\defcitealias{Gratton10_masslossGCs}{G10}
\defcitealias{Howell22_M4}{H22}

\title[A Seismic Mass Study of M80]{First Asteroseismic Analysis of the Globular Cluster M80: Multiple Populations and Stellar Mass Loss}

\author[M. Howell et al.]{
Madeline Howell,$^{1,2}$\thanks{E-mail: madeline.howell1@monash.edu}
Simon W. Campbell,$^{1,2}$
Dennis Stello$^{3,4,5,2}$
and Gayandhi M. De Silva$^{6,7,2}$
\\
$^{1}$School of Physics and Astronomy, Monash University, Clayton, VIC 3800, Australia\\
$^{2}$ARC Centre of Excellence for Astrophysics in Three Dimensions (ASTRO-3D), Australia\\
$^{3}$School of Physics, University of New South Wales, NSW 2052, Sydney, Australia\\
$^{4}$Sydney Institute for Astronomy (SIfA), School of Physics, University of Sydney, NSW 2006, Sydney, Australia\\
$^{5}$Stellar Astrophysics Centre, Department of Physics and Astronomy, Aarhus University, DK-8000 Aarhus C, Denmark\\
$^{6}$Australian Astronomical Optics, Faculty of Science and Engineering, Macquarie University, Macquarie Park, NSW 2113, Australia\\
$^{7}$School of Mathematical and Physical Sciences, Faculty of Science and Engineering, Macquarie University, Macquarie Park, NSW 2113, Australia\\
}

\date{Accepted XXX. Received YYY; in original form ZZZ}

\pubyear{2023}

\begin{document}
\label{firstpage}
\pagerange{\pageref{firstpage}--\pageref{lastpage}}
\maketitle

\begin{abstract}
Asteroseismology provides a new avenue for accurately measuring the masses of evolved globular cluster (GC) stars through the detection of their solar-like oscillations. We present the first detections of solar-like oscillations in 47 red giant branch (RGB) and early asymptotic giant branch (EAGB) stars in the metal-poor GC M80; only the second ever with measured seismic masses. We investigate two major areas of stellar evolution and GC science; the multiple populations and stellar mass-loss. We detected a distinct bimodality in the EAGB mass distribution. We showed that this is likely due to sub-population membership. If confirmed, it would be the first direct measurement of a mass difference between sub-populations. A mass difference was not detected between the sub-populations in our RGB sample. We instead measured an average RGB mass of $0.782\pm0.009~\msun$, which we interpret as the average between the sub-populations. Differing mass-loss rates on the RGB has been proposed as the second parameter that could explain the horizontal branch (HB) morphology variations between GCs. We calculated an integrated RGB mass-loss separately for each sub-population: $0.12\pm0.02~\msun$ (SP1) and $0.25\pm0.02~\msun$ (SP2). Thus, SP2 stars have greatly enhanced mass-loss on the RGB. Mass-loss is thought to scale with metallicity, which we confirm by comparing our results to a higher metallicity GC, M4. We also find that M80 stars have insignificant mass-loss on the HB. This is different to M4, suggesting that there is a metallicity and temperature dependence in the HB mass-loss. Finally, our study shows the robustness of the $\Delta\nu$-independent mass scaling relation in the low-metallicity (and low-surface gravity) regime. 
\end{abstract}


\begin{keywords}
asteroseismology – stars: low-mass – stars: mass-loss – stars: oscillations; galaxies: star clusters: individual: NGC 6093 (M80)
\end{keywords}



\section{Introduction}
\label{sec:intro}


With the advent of the \textit{Kepler} telescope \citep{Koch10_kepler} and the \textit{K2} mission \citep{Howell14_K2_mission}, high-quality long-time series photometry now exists for globular clusters (GC), providing us with the opportunity to study stellar evolution with increased accuracy using asteroseismology. Since the release of the \textit{K2} photometry, there are reported detections of solar-like oscillations in only one GC, M4 \citep{Miglio_M4_study,Tailo22_M4,Howell22_M4}. GCs are important objects for studying stellar evolution in the low mass and low-metallicity regime, and hence it is essential that we increase the number of GCs that have asteroseismic measurements. Here we present the first detections of solar-like oscillations in the metal-poor ([Fe/H]~$\sim -1.8$; \citealt{Caretta15_M80}) globular cluster, M80; only the second GC observed in the \textit{K2} mission with detections of solar-like oscillations. 

GCs contain essentially co-eval and co-spatial stellar populations with homogeneous heavy element distributions, and have been described as \textit{Astronomers' laboratories} \citep{Moehler02_astronmers'_labs}. These characteristics make them excellent objects for studying stellar evolution -- they represent one of the most robust ways in which we can investigate highly uncertain aspects of how stars evolve, such as mass loss. In this paper, we focus on inferring the stellar mass loss in the GC M80 through direct asteroseismic mass determinations. The precision of the seismic mass measurements also allows us to investigate the multiple population phenomenon in GCs.

Mass loss currently remains a major uncertainty in stellar modelling. For mass loss on the red giant branch (RGB), empirical relations such as the Reimers' scheme \citep{Reimers1975_massloss_rate} or the Schr\"oder \& Cuntz scheme \citep{Schroder_cunts05_massloss_rate} are used in models, which are dependent on bulk stellar properties (e.g. luminosity ($\mathrm{L}$), radius ($\mathrm{R}$), mass ($\mathrm{M}$), surface gravity ($\log\mathrm{g}$), and effective temperature ($T_{\text{eff}}$)). These empirical schemes are reliant on a scaling parameter, $\eta$, which needs to be calibrated against observations. This scaling parameter introduces uncertainty into the mass loss rates and will impact the evolution of the RGB, and also subsequent phases such as the horizontal branch (HB), asymptotic giant branch (AGB), and the final stellar remnant. To reduce the uncertainty in this free parameter, an integrated mass loss can be quantified by taking the differences in the averaged masses of various evolutionary phases \citep[e.g.][]{Miglio12_OCstudy, Tailo22_M4, Howell22_M4}.

Stellar mass loss is thought to be metallicity dependent, where high-metallicity stars lose more mass than low-metallicity stars. However, a mass loss-metallicity trend is not usually explicitly\footnote{Some mass loss variation does come about indirectly through the models. For example at low metallicity the stellar structure changes (lower R, higher $T_{\text{eff}}$).}
accounted for in stellar modelling. GCs serve as unique testbeds to estimate the amount of mass lost in the RGB evolutionary phase for a range of low metallicities. We can also use them to derive the relationship between the mass loss on the RGB and metallicity, which has been attempted in the literature. For example, \citet[][hereby \citetalias{Gratton10_masslossGCs}]{Gratton10_masslossGCs} and \citet[][hereby \citetalias{Tailo20_massloss_difference_multipops}]{Tailo20_massloss_difference_multipops} determined a mass loss-metallicity trend using an indirect method to estimate masses of HB stars from models that match photometric colour-magnitude diagrams. \citet[][hereby \citetalias{Origlia14_masslossGCs}]{Origlia14_masslossGCs} also investigated the mass loss-metallicity relationship by using infrared colour excesses of circumstellar disks around dusty RGB stars to estimate the mass loss. We note that the existence of significant dust production in low mass and low luminosity RGB stars has been questioned (e.g. \citealt{Boyer10_nodustproduction1,McDonald11_nodustproduction2}). Each study estimated a similar linear trend between the metallicity and the mass lost on the RGB, albeit with an offset of $\sim0.1~\msun$. To better determine this trend, we need more accurate measurements of mass loss at different metallicities. Asteroseismic masses -- which do not rely on measurements of dust or indirect mass measurements involving models -- remains the most robust way of measuring an integrated mass loss (e.g. \citealt{Howell22_M4,Tailo22_M4}), and testing the stellar mass loss-metallicity dependence.


We can accurately estimate the asteroseismic masses of individual stars from their solar-like oscillations (stochastically-excited pressure-mode oscillations). These asteroseismic signals are characterised by two global quantities, which are correlated to stellar properties; $\nu_{\text{max}}$ is related to the surface gravity and temperature: $\nu_{\text{max}} \propto gT_{\text{eff}}^{-1/2}$, and $\Delta\nu$ is related to the mean density of the star: $\Delta\nu \propto \overline{\rho}^{1/2}$ \citep{Ulrich1986_scaling_relation1,Brown1991_scaling_relation2,Kjeldsen1995_scaling_relation3}. From these relations, combined with the Stefan-Boltzmann luminosity law ($L\propto R^2T_{\text{eff}}^4$), four seismic masses can be calculated for each star from the following set of equations:
\begin{align}
\label{eq:mass_relation}
    &\left(\frac{M}{M_{\odot}}\right)\simeq\left(\frac{\nu_{\text{max}}}{\nu_{\text{max},\odot}}\right)^3\left(\frac{\Delta\nu}{\Delta\nu_{\odot}}\right)^{-4}\left(\frac{T_{\text{eff}}}{T_{\text{eff},\odot}}\right)^{3/2}\\
\label{eq:mass_relation2}
    &\left(\frac{M}{M_{\odot}}\right)\simeq\left(\frac{\Delta\nu}{\Delta\nu_{\odot}}\right)^{2}\left(\frac{L}{L_{\odot}}\right)^{3/2}\left(\frac{T_{\text{eff}}}{T_{\text{eff},\odot}}\right)^{-6}\\
\label{eq:mass_relation3}
    &\left(\frac{M}{M_{\odot}}\right)\simeq\left(\frac{\nu_{\text{max}}}{\nu_{\text{max},\odot}}\right)\left(\frac{L}{L_{\odot}}\right)\left(\frac{T_{\text{eff}}}{T_{\text{eff},\odot}}\right)^{-7/2}\\
\label{eq:mass_relation4}
    &\left(\frac{M}{M_{\odot}}\right)\simeq\left(\frac{\nu_{\text{max}}}{\nu_{\text{max},\odot}}\right)^{12/5}\left(\frac{\Delta\nu}{\Delta\nu_{\odot}}\right)^{-14/5}\left(\frac{L}{L_{\odot}}\right)^{3/10}
\end{align}

Asteroseismic masses have been successfully used to measure mass loss in open clusters (e.g. NGC6791, NGC6819, and M67; \citealt{Miglio12_OCstudy,Handberg17_OC_study,Stello16_M67}), where they found small mass differences between the RGB and core-He burning phases of $-0.03 \leq \Delta M \leq 0.09~\msun$. This small mass loss is likely due to the young ages of these clusters, and agrees with cluster models using a Reimers’ mass loss formulation (\citealt{Miglio12_OCstudy}). In contrast, GC stars are older, and are expected to present larger mass differences between the RGB and HB ($\sim0.2~\msun$; e.g. \citealt{McDonald11_omega_centauri_massloss,Salaris16}).

Detections of solar-like oscillations in GCs were initially attempted using ground-based \citep{Frandsen07_M4asteroseismology} and \textit{Hubble Space Telescope} photometry \citep{Edmonds96_Tuc47asteroseismology, Stello09_HST_GC_seismo}, however no useful measurements were derived for the global asteroseismic quantities. Unambiguous measurements of $\Delta\nu$ and $\nu_{\text{max}}$ in GCs stars were first detected by \citet{Miglio_M4_study} in the cluster M4. They measured masses for seven RGB stars, and one red HB (RHB) star, however they weren't able to estimate an integrated RGB mass loss due to the small number of stars. Recently, \citet[][hereby \citetalias{Tailo22_M4}]{Tailo22_M4} increased the sample of M4 stars with detected solar-like oscillations to 37 stars:  32 RGB stars and 6 RHB stars. They calculated a RGB integrated mass loss of $\Delta \overline{M} = 0.227\pm0.028~\msun$. Finally, \citet[][hereby \citetalias{Howell22_M4}]{Howell22_M4} reported the largest seismic sample yet with 75 detections of solar-like oscillations in red giants: 59 RGB, 11 RHB, and 5 EAGB. They reported an integrated RGB mass loss of $\Delta \overline{M} = 0.17\pm0.01~\msun$, significantly lower than the \citetalias{Tailo22_M4} value. We suggest that the different methods of calculating temperatures -- an iterative photometric process in \citetalias{Tailo22_M4}, and spectroscopic and scaled photometric temperatures in \citetalias{Howell22_M4} -- could be the fundamental reason for the discrepancy in the RGB integrated mass loss result.

We can also extend this analysis of the asteroseismic masses to study the multiple populations in GCs. Multiple populations have been observed in a large number of GCS; e.g. NGC 6397 and NGC 6752 \citep{Gratton01_Mulpop_obs1}, 47 Tuc \citep{Anderson09_Mulpop_obs2}, M80 \citep{Caretta15_M80}. It is thought that all GCs contain typically two -- sometimes more -- sub-populations which vary in light elemental abundances \citep[e.g. C, N, O, Na, O, and He;][]{Sneden1999_light_elemental_abundances1,Gratton12_light_elemental_abundances2}. The variation in helium, which is a key element for stellar evolution, will result in different current observed stellar masses between the populations \citep{MacLean18_chloes_paper,Jang19_2pops_models}, assuming the GC stellar populations are essentially co-eval. The He-normal population are expected to have higher stellar masses, and the He-rich population will have lower masses. 

To determine if there is a mass difference between sub-populations in the GC M4, \citetalias{Howell22_M4} searched the distribution of RGB asteroseismic masses. They showed a mass dispersion in the RGB sample that was higher than expected, consistent with a hidden mass variation. This broad distribution agreed with the expected sub-population masses from detailed models for the two chemical sub-populations in M4 \citep{MacLean18_chloes_paper}. This broad mass distribution could be due to multiple populations. Chemical abundances are needed to confirm or deny the conclusions in \citetalias{Howell22_M4}. \citetalias{Tailo22_M4} also investigated the multiple populations with their asteroseismic masses in combination with sub-population classification from the $C_{\text{U,B,I}}$ index (see \citealt{Milone13_Cubi,Monelli13_Cubi} for details). They similarly found no significant evidence of a mass difference between populations. The mass difference between sub-populations in M4 is thought to be $\sim0.017-0.04~\msun$ \citep{MacLean18_chloes_paper,Tailo19_M4_multipop_massloss}, which is within the mass uncertainties of both studies. Hence, it may not be possible to detect a strong bimodality with the currently available photometry for the cluster M4. However for the cluster M80, it has been proposed that the peculiar kinematics and radial distributions of the multiple populations indicates a mass difference between the sub-populations of $0.05-0.10~\msun$ \citep{Dalessandra18_multi_pop_M801}. This larger mass separation of M80's sub-populations could be easier to detect in each evolutionary phase, and hence we may be able to observe a clear bimodal mass distribution for this cluster.

An asteroseismic study of multiple populations also provides the opportunity to determine if there are integrated mass loss differences between the sub-populations in GCs, which could inform the second parameter problem. Metallicity is known to have significant impacts on the effective temperatures of HB stars, however it cannot fully describe the variation in HB morphology observed between GCs \citep[e.g.][]{VanDenBerg1967_2ndparamproblem,Sandage1967_2ndparam,FusiPecci1993_2ndparamproblem}. Hence to describe this variation, additional parameter(s) that influence the HB temperatures are needed. It has been proposed that He enhancement and differing RGB mass loss rates are both major factors in determining the temperatures of HB stars in GCs (e.g. \citealt{Tailo19_M4_multipop_massloss}, \citetalias{Tailo20_massloss_difference_multipops}). If we do detect bimodal mass distributions in our sample, we could quantify the integrated mass loss differences between the sub-populations. Given that there are known He variations between the sub-populations, this could assist in investigating whether He enhancement and differing RGB mass loss rates are indeed the second parameters that define the HB morphology.

In this study, we report the first asteroseismic detections in the GC M80\footnote{We note that \textit{K2} observations have been used study variable stars in this cluster (\citealt{Plachy_M80_K2}; Moln\'{a}r et al., in prep), but no solar-like oscillations have been reported in the literature so far.}. The GC M80 is fainter compared to M4, and solar-like oscillations are not detectable at the HB magnitude ($G_{\text{mag}} \sim 16$) with \textit{K2} photometry. Therefore, we cannot estimate a seismic RGB integrated mass loss by finding the differences between the RGB and HB masses. We can instead calculate a seismic mass difference between the RGB and early AGB (EAGB; the phase of evolution directly after the HB), which we refer to as $\Delta M_{RGB-EAGB}$. In Section~\ref{sec:data_prep}, we describe our methodology of target selections, and the detrending of \textit{K2} photometric data. In Section~\ref{sec:seismic_quantities} we outline our method of measuring the global asteroseismic parameters. Section~\ref{sec:stellar_params} discusses how we estimated the stellar properties $T_{\text{eff}}$, luminosity, and radius. We present the seismic mass distributions for our sample, an analysis of the multiple populations, and the integrated mass loss estimate in Section~\ref{sec:mass_results}. We also show a comparison of our mass estimates to isochrones. Finally, in Section~\ref{sec:discussion} we discuss the implications of the mass results and the integrated mass loss $\Delta M_{RGB-EAGB}$ for M80, and compare them to M4 mass results from previous literature.

\section{Data Preparation and Analysis}
\label{sec:data_prep}

\subsection{Target sample selection}
\label{sec:target_selection}

\begin{figure*}
	\centering
	\includegraphics[width=2\columnwidth]{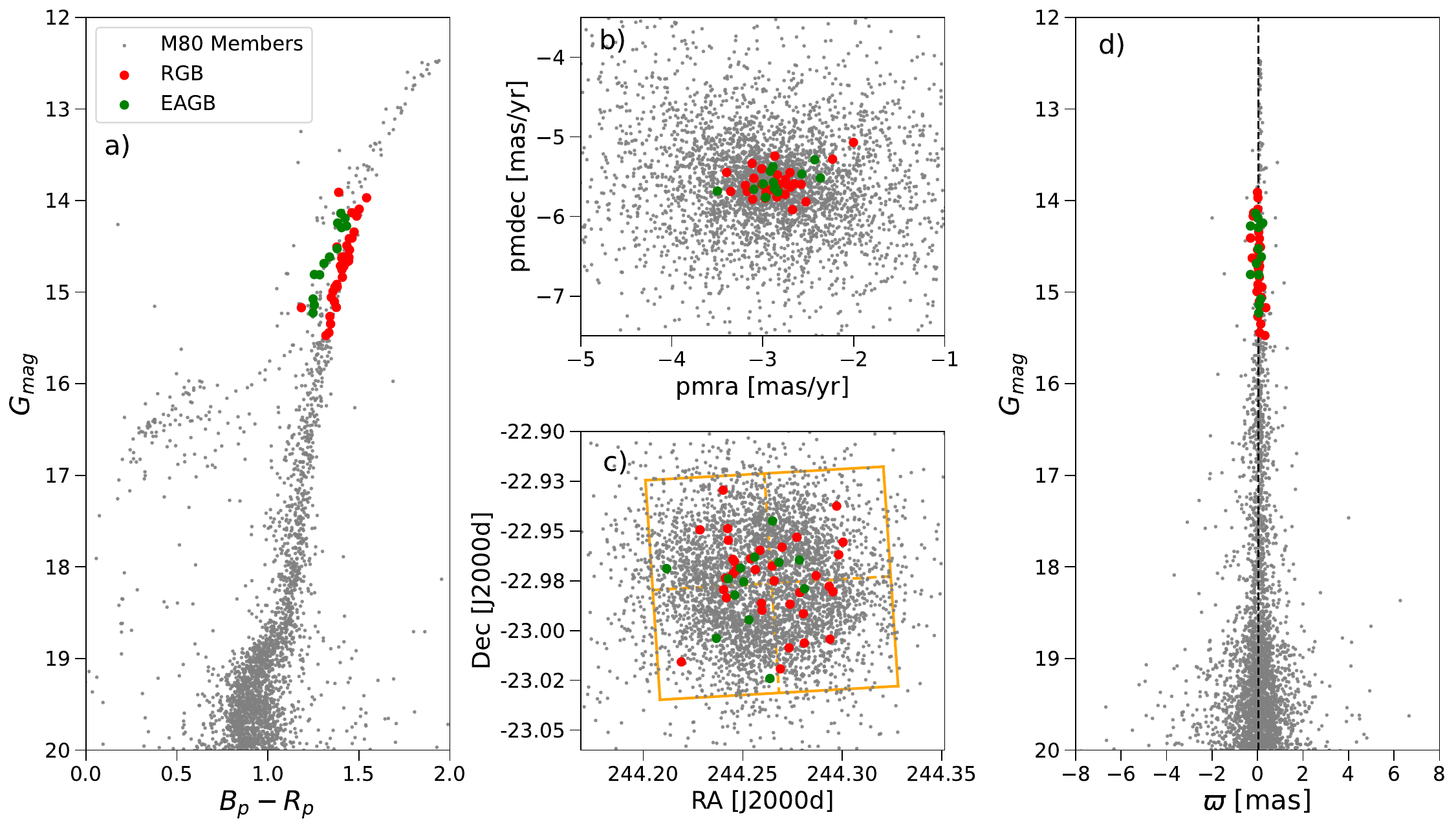}
	\caption{\textbf{(a) } Colour-magnitude diagram (CMD) of our M80 sample (coloured points) and a \textit{Gaia} EDR3 membership sample \citep[grey;][]{Vasiliev21_GalacticGC_memberships}. Stars for which we were able to detect solar-like oscillations are classified into two evolutionary phases; RGB and EAGB, and are indicated by the larger coloured points. \textbf{(b) } \textit{Gaia} proper motions of the same sample. \textbf{(c) } The spatial positions of the sample. We also show the M80 \textit{K2} superstamp (orange). \textbf{(d)} The \textit{Gaia} EDR3 parallaxes, $\varpi$, for the cluster members and our seismic sample. The average parallax for the cluster is indicated by the dashed line.}
	\label{fig:CMD}
\end{figure*}

We started with the membership survey from \citet{Vasiliev21_GalacticGC_memberships}. They used the \textit{Gaia} EDR3 proper motions and parallaxes (see panels (b) and (d) in Fig.~\ref{fig:CMD}) to assign a membership probability metric. To be included in our member sample, a star needed a membership probability $\geq 90\%$. Based on the \citetalias{Howell22_M4} results, we set a cut-off magnitude in the \textit{Gaia} G-band of $15.5$~mag. This ensured that an asteroseismic signal could be detected above the \textit{K2} noise level, since the amplitudes of the solar-like oscillations decrease as luminosity decreases \citep{Stello2011_AmpSLO_scaling}.

\begin{figure*}
	\centering
	\includegraphics[width=2\columnwidth]{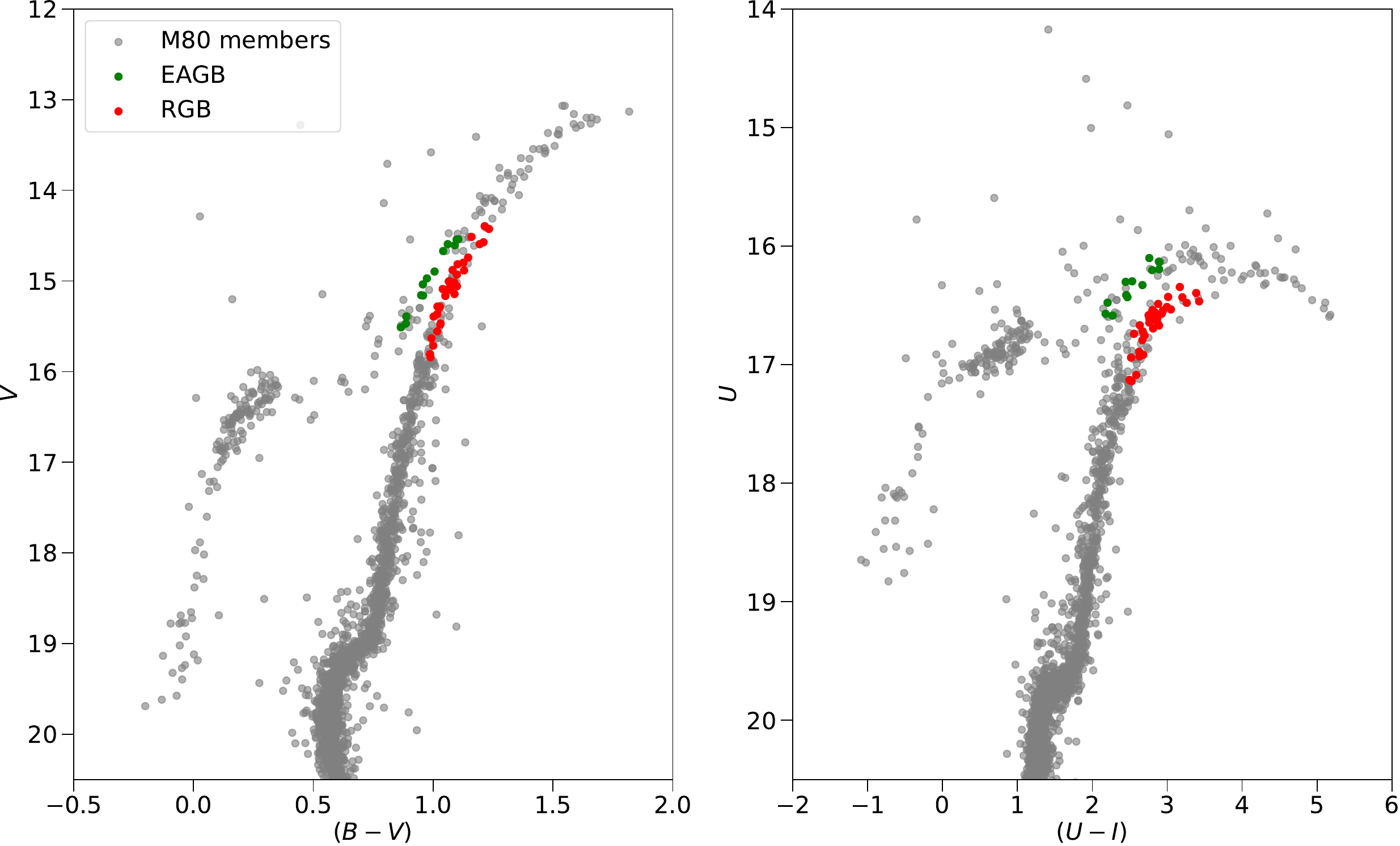}
	\caption{The V,(B-V) and U,(U-I) CMDs for our M80 sample (coloured points), overplotted onto the full M80 sample from the \citet{Stetson19_GCS_UBVRI_photometry} photometric sample (grey points). These CMDs were used to confirm the evolutionary phases of the stars.}
	\label{fig:UBVI_CMD}
\end{figure*}

We passed this membership sample through the \texttt{K2fov} algorithm \citep{K2FOV} to identify which stars were on silicon in the \textit{K2} observations (Fig.~\ref{fig:CMD}(c)). This reduced sample was then classified into evolutionary phases, using UBVRI photometry from \citet{Stetson19_GCS_UBVRI_photometry}. It has been shown that the RGB and AGB clearly separate in the B-V and U-I colours (e.g. \citealt{MacLean18_chloes_paper}), which was similarly observed for our M80 sample (Fig.~\ref{fig:UBVI_CMD}). Lastly, we assigned our own star identifications using the same naming convention as \citetalias{Howell22_M4}.

\subsection{Aperture masks \& the photometric detrending pipeline}
\label{sec:lc_detrending}
M80 was observed in Campaign 2 of the \textit{K2} mission, where the cluster centre was captured in 4 target pixel files: EPIC200004386, EPIC200004387, EPIC200004388 \& EPIC200004389 (Fig.~\ref{fig:CMD}(c)). Each target pixel file consisted of a grid of $50\times50$ pixels, and comprised of $\sim80$ days of observations across 3856 30-minute long cadences. In this study, we used the same method as \citetalias{Howell22_M4} to produce the optimal aperture masks and detrended \textit{K2} light curves using \texttt{Lightkurve} \citep{Lightkurve}. 

In summary, aperture masks for each star were initially found using an automated method which includes pixels that were within a certain threshold level of the brightest pixel (assumed to be around the centre of the star). This method also ensures that pixels from neighbouring stars are mostly excluded, however it was difficult to determine at this stage if there is photometric contamination between nearby stars (see Section~\ref{sec:seismic_quantities} for further analysis). We finalised the individual aperture masks with a manual check, which involved finding the optimal pixel mask by visually inspecting the resultant power spectrum. 

The detrending pipeline was a two step process: first applying a pixel-level decorrelator to the target pixel file (based on the method by \citealt{Deming15_PLD_creator}) to remove systematic effects from individual pixels (e.g. instrumental drifts), and then removing long-term trends from the photometric light curves by using a self-flat fielding corrector \citep{Vanderburg14_SFF_cleaning}. Refer to Section 2.2 of \citetalias{Howell22_M4} for a detailed description of the methodology. 

Comparing the $\nu_{\text{max}}$ of stars at a similar location on the RGB and AGB as the \citetalias{Howell22_M4} M4 sample, we expected that the solar-like oscillations for our entire M80 sample should be in the frequency region $\leq 20~\mu\mathrm{Hz}$. This frequency regime in the spectra was manually searched for potential solar-like oscillations. This resulted in a sample of 49 red giants with detected solar-like oscillations: 35 RGB stars and 14 EAGB stars. 


\section{Global Asteroseismic Quantities}
\label{sec:seismic_quantities}

The global asteroseismic quantities, $\nu_{\text{max}}$ and $\Delta\nu$, were estimated from our detrended light curves using the \texttt{pySYD} pipeline \citep{pySYD_chontos}, and the method detailed in the \citetalias{Howell22_M4} study. Following the \citetalias{Howell22_M4} findings for the fitting of the background signal of bright stars, we used a linear background model for the power excess envelope. This made the code more stable for noisy spectra, when attempting the Monte Carlo sampling routine for the uncertainty estimation. We adopted a quality flag system; marginal detection (MD) and detection (D). This assignment was based on the visual inspection of each power spectrum. If a star's power excess was observed to be `noisy', it was labelled as a `MD'. Otherwise, a `D' flag was assigned.

The higher stellar density of M80 increased the probability that a star's power spectrum could be photometrically contaminated by a neighbouring star. To test if there was contamination in a star's spectrum, we used the `confusion from blends' metric introduced in \citet{Stello22_tess_red_giants} for TESS red giants. This method uses the Shape-Base Distance (SBD) algorithm \citep{Paparrizos_2016_kshapes} to quantify the similarity between two power spectra. A perfect correlation in frequency space corresponds to a metric of 0, and an anticorrelation corresponds to a metric of 2. Potentially contaminated stars were identified if they had aperture masks within 2 pixels (angular distance of $\sim8''$) of a neighbouring star. The SBD metric was calculated for each potentially contaminated star across the frequency range of the FWHM of the oscillation power excess (defined as $\delta\nu_{\text{FWHM}}=0.59\nu_{\text{max}}^{0.9}$; \citealt{Mosser2010}). 

First, we tested for the range of SBD metric values for our data. Two tests were completed; a similarity test to find the SBD metric between the same power spectrum, and a dissimilarity test to find the metric of two power spectra which look very different. For the similarity test, we reached a SBD metric of $10^{-16}$ matching the expected lower limit of zero. For the dissimilarity value test, we compared three pairs of different spectra, and the maximum SBD metric we obtained was $\sim0.7$, representing an approximate upper limit for our sample. 

\begin{figure}
	\centering
	\includegraphics[width=\columnwidth]{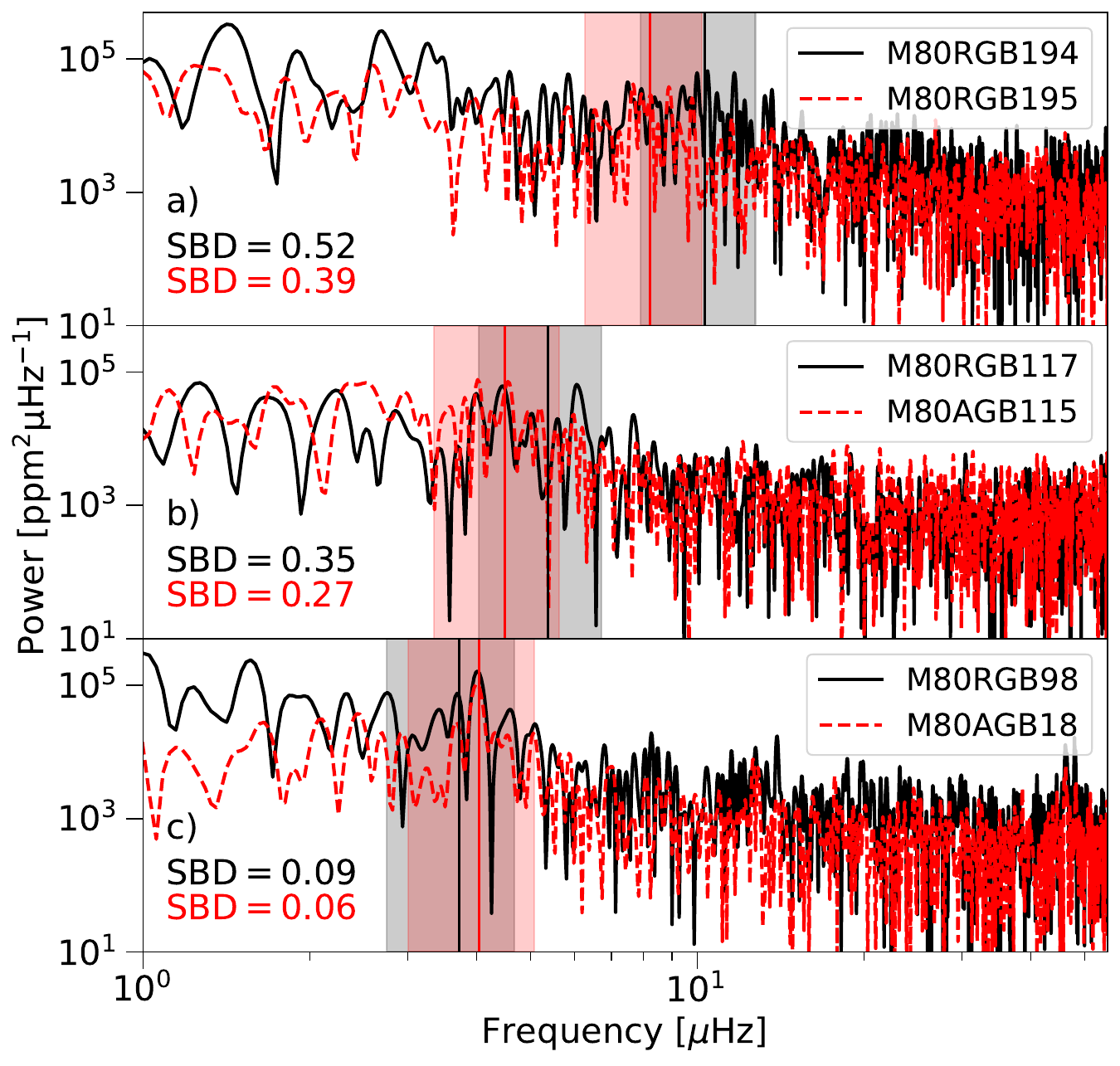}
	\caption{Each panel shows the power spectra for two stars which are spatially close to each other and have similar $\nu_{\text{max}}$ values. We provide the SBD metric (colours correspond to the power spectra colours) for each star's envelope, where values close to zero indicate that the spectra are similar, and values closer to $\sim 0.7$ are dissimilar. The vertical lines correspond to the stars' measured $\nu_{\text{max}}$ values, and the shaded region represents the FWHM envelope for each star's solar-like oscillations. Panel (a) shows a pair of stars which have dissimilar power spectra, panel (b) shows an ambiguous case, and the stars in panel (c) were found to be very similar.}
	\label{fig:SBD_spectra}
\end{figure}

We demonstrate some examples of the SBD method in Figure~\ref{fig:SBD_spectra}, where the power spectra for three pairs of neighbouring stars with similar measured $\nu_{\text{max}}$ are shown. The calculated SBD metrics for each star are also included in this figure. Panel (a) shows an example of stars with a large SBD metric suggesting that we are detecting separate signals. Panel (b) of Figure~\ref{fig:SBD_spectra} shows an ambiguous case, with a moderately high SBD metric, but with similar solar-like oscillations upon visual inspection. These stars were consequently assigned the quality flag `MD'. 

The pair of stars, M80RGB98 and M80AGB18 (shown in Fig.~\ref{fig:SBD_spectra}(c)), were found to have a SBD metric close to zero (SBD~$\sim0.06-0.09$) in their identified oscillation envelope, indicating that we could be detecting the same signal in both spectra. Comparing their magnitudes, M80RGB98 is bright in the \textit{Gaia} G-band with a magnitude of 13.35 mag, which should correspond to a $\nu_{\text{max}}<1~\mu\mathrm{Hz}$ (based on the results in the \citetalias{Howell22_M4} study). For this reason, we concluded that the signal we are observing most likely belongs to the star M80AGB18, because we do not expect to see solar-like oscillations for the star M80RGB98 with the resolution of the \textit{K2} data. This was also found for another star with a similar bright magnitude, M80AGB55. Both of these bright stars were removed from our sample. This reduced our final seismic sample to 47 red giants (Fig.~\ref{fig:CMD}). A total of 19 stars have flag MD and 28 stars have flag D.

\begin{figure}
	\centering
	\includegraphics[width=\columnwidth]{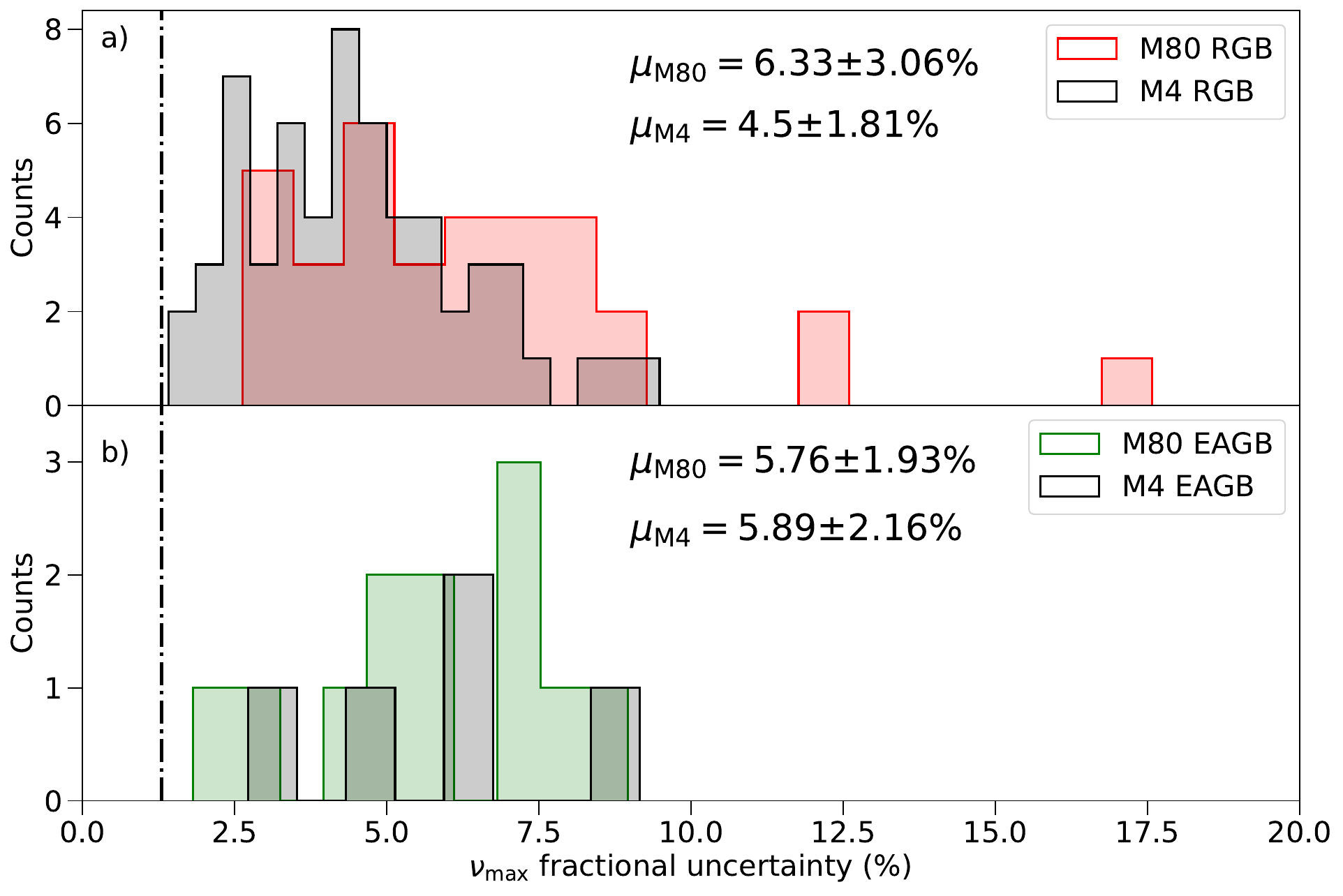}
	\caption{\textbf{(a)} Histogram showing the distribution of fractional (\%) uncertainties from the $\nu_{\text{max}}$ estimations for our RGB sample. Our percentage error distributions are compared to the error distributions from the \citetalias{Howell22_M4} M4 sample (black) and also the median fractional uncertainties from K2GAP for RGB/EAGB stars (dashed black line). Our median fractional uncertainties, $\mu$, and the $1\sigma$ scatter are also written on the plot for each evolutionary phase. \textbf{(b)} The same as the top panel but for our EAGB sample. }
	\label{fig:percentage_errors}
\end{figure}

The fractional uncertainty distributions for $\nu_{\text{max}}$ were compared to the \citetalias{Howell22_M4} M4 results, and to the K2 Galactic Archaeology Program data release 3 (K2GAP; \citealt{Zinn21_K2GAP_DR3}) in Figure~\ref{fig:percentage_errors}. Our M80 $\nu_{\text{max}}$ uncertainties are broadly consistent with the uncertainty distribution of M4, and like for M4, larger than the K2GAP results.

As stated in the \citetalias{Howell22_M4} study, we would like to emphasise that the $\Delta\nu$ measurements for M80 stars from the \textit{K2} photometry should be used with caution. This is due to the low signal-to-noise of our sample's power excesses, and also the relatively low resolution of the \textit{K2} data, where we cannot observe the solar-like oscillations of stars with $\nu_{\text{max}} \leq 2~\mu\mathrm{Hz}$. This is further exacerbated by the inherent low number of observable modes at low $\nu_{\text{max}}$ values \citep{Stello2014_reduced_modes_RGBtip}, which makes it difficult to identify the repeating modal pattern needed to measure an accurate $\Delta\nu$. Due to these reasons, we do not trust the formal uncertainties on $\Delta\nu$, and hence did not compare them to previous studies. 

The final global asteroseismic quantities are provided in Table~\ref{tab:final_results}.

\section{Stellar Parameters}
\label{sec:stellar_params}
\subsection{Effective temperatures}
\label{sec:teffs_dust}
The effective temperatures of our stellar sample were calculated using photometric 2MASS $JHK$ \citep{2MASS} and Johnson-Cousins $UBVRI$ magnitudes \citep{Stetson19_GCS_UBVRI_photometry}. We followed the same method as \citetalias{Howell22_M4}, where Equation 8 from \citet{Fitzpatrick07_ev_k} was implemented to find the dereddened $(V-K)$ colour. Photometric temperatures were found using the colour-$T_{\text{eff}}$ relation for giant stars from \citet{Gonzales_Hernandez_colour_teff_relation}, using a metallicity of [Fe/H]~$ =-1.791\pm0.006\text{(rand)}\pm0.076\text{(sys)}$ \citep{Caretta15_M80}. Previous M80 studies have used a constant dust extinction of $E(B-V)=0.18$ \citep{Harris2010,Caretta15_M80} for reddening corrections to photometry, which we adopt for this study. However, this doesn't account for local dust variations, which can introduce systematic and random errors into the effective temperatures and luminosities.    

As a cross-check, we calculated another set of photometric temperatures using the \textit{Gaia} colour ($B_P-R_P$). We used the $T_{\text{eff}}$-colour relation for \textit{Gaia} pass-band photometry from \citet{Mucciarelli20_gaia_teff_colour_relation}, where the dust-corrected \textit{Gaia} colour was $(B_{P}-R_{P})_0 = (B_P-R_P) - E(B_P-R_P)$. We assumed a constant extinction in our entire sample of $E(B_P-R_P) = 0.52$, which was determined by finding the mean value provided by the \textit{Gaia} DR3 catalog for a M80 membership sample from \citet{Vasiliev21_GalacticGC_memberships} (shown as grey dots in Fig.~\ref{fig:CMD}). We found that the effective temperatures for our seismic sample from the $(V-K)$ and $(B_P-R_P)$ colours were similar, and had an average offset of $+8.5~\mathrm{K}$. From the scatter in the $T_{\text{eff}}$ residuals of the two photometric scales, we estimated an uncertainty for the temperatures of $\pm110~\mathrm{K}$.

Within the sample, there were three outliers with a $T_{\text{eff}}$ difference larger than $300~\mathrm{K}$, and the photometric errors of each were individually inspected. Two of the outliers, M80AGB84 and M80RGB09, had an `E' quality flag in the 2MASS $K$ magnitude. This flag indicates poor photometry, and we instead assigned the \textit{Gaia}-based $T_{\text{eff}}$ for these stars. This was also done for two other stars which had the 2MASS `E' quality flag for the $K_{\text{mag}}$. In addition, we checked the \textit{Gaia} photometric errors on the $B_P$ and $R_P$ magnitudes and the RUWE metric for the stars M80AGB84 and M80RGB09, which were found to be consistent with the cluster's average values. The last outlier star in the $T_{\text{eff}}$ differences, M80RGB203, had no indication of bad photometry in the $V$ or $K$ magnitudes, but was found to have an uncertainty for the $B_P$ magnitude which was an order of magnitude larger than the average error for the cluster. Hence, it was assumed that for this star the ($V-K$) $T_{\text{eff}}$ was the more reliable value.

As there is no dust survey for M80, we are unable to test for possible systematics in the photometric $T_{\text{eff}}$ due to differential reddening, or know about the properties of the dust. This includes the spatial variation in the dust, or the dust grain type which is related to the parameter $R_V$; $R_{V} = A_V/E(B-V)$, where $A_V$ is the total extinction in the visual band. The \citetalias{Howell22_M4} study showed that even with a boutique dust survey of the M4 cluster, extinction still remains a significant uncertainty in the photometric temperatures, and hence the seismic masses. Due to this, we ideally want to remove the dependence of dust reddening, which can be accomplished with extinction-independent spectroscopic temperatures in the future. Our final photometric $T_{\text{eff}}$ values are provided in Appendix~\ref{tab:final_results}.

\subsection{Bolometric luminosity}
\label{sec:luminosity}
The luminosities were calculated using the same method as the \citetalias{Howell22_M4} study:
\begin{equation}
\label{eq:phot_l}
    \log(L/L_{\odot}) = -0.4\left[V_0-(m-M)_0+BC-M_{\text{bol},\odot}\right]
\end{equation}
and using the bolometric correction ($BC$) for giant stars \citep[Eq. 18 from][]{Alonso99_bolometric_correction}. The value for the bolometric magnitude of the Sun is found in Table~\ref{tab:solar_values}. For the true distance modulus for M80, we used $(m-M)_0 = 15.06\pm0.07$, which was calculated from the literature search in \citet{Baumgardt21_GCs_membership}\footnote{We calculated our own average distance modulus which excluded the distance moduli estimates that had large uncertainties or were more than $1\sigma$ away from the average. The uncertainty represents the dispersion in the estimates.}. The dereddened visual band magnitude, $V_0$, was calculated with $V_0 = V - A_V$, where the typical extinction ratio for the Galaxy, $R_V = 3.1$, was used. We note that $R_V$ is not constant for the entire galaxy, for example M4 was found to have a value of $R_V = 3.62\pm0.07$ \citep{Hendricks12_M4_dustmap}. As there are no public extinction surveys of M80, we are unsure whether using $R_V = 3.1$ is appropriate for this cluster. If not, it could systematically shift the luminosities, and hence offset the final seismic masses. In Section~\ref{sec:mass_systematics_discussion}, we test the systematic effect of extinction on the seismic masses, which is related to the systematic shift from the extinction ratio, because $R_V$ is dependent on $E(B-V)$. The bolometric luminosities for our seismic sample can be found in Table~\ref{tab:final_results}.

\begin{table}
\centering
\footnotesize
\caption{The solar reference values used in this study. The solar asteroseismic quantities are determined from the \texttt{SYD} pipeline.}
    \begin{tabular}{lcc}
      \hline 
      Parameter & Value & Reference\\
      \hline 
      $M_{\text{bol},\odot}$ & $4.74$ & \citet{Mamajek15_mbol}\\
      $\nu_{\text{max},\odot}$ & $3090\pm30~\mu\mathrm{Hz}$ & \citet{Huber11_solar_syd_values}\\
      $\Delta\nu_{\odot}$ & $135.1\pm0.1~\mu\mathrm{Hz}$ & \citet{Huber11_solar_syd_values} \\
      $T_{\text{eff},\odot}$ & $5772\pm0.8~\mathrm{K}$ & \citet{Mamajek15_B3} \\
      \hline 
    \end{tabular}
\label{tab:solar_values}
\end{table}

\subsection{Radii \& seismic surface gravity}
\label{sec:radii}

Individual seismic radii were calculated for each star using:
\begin{equation}
\label{eq:radius_relation}
    \left(\frac{R}{R_{\odot}}\right)\simeq\left(\frac{\nu_{\text{max}}}{\nu_{\text{max},\odot}}\right)\left(\frac{\Delta\nu}{\Delta\nu_{\odot}}\right)^{-2}\left(\frac{T_{\text{eff}}}{T_{\text{eff},\odot}}\right)^{1/2}
\end{equation}
where the solar reference values can be found in Table~\ref{tab:solar_values}. We provide the seismic radii estimates for our seismic sample in Table~\ref{tab:final_results}. 

\begin{figure}
	\centering
	\includegraphics[width=\columnwidth]{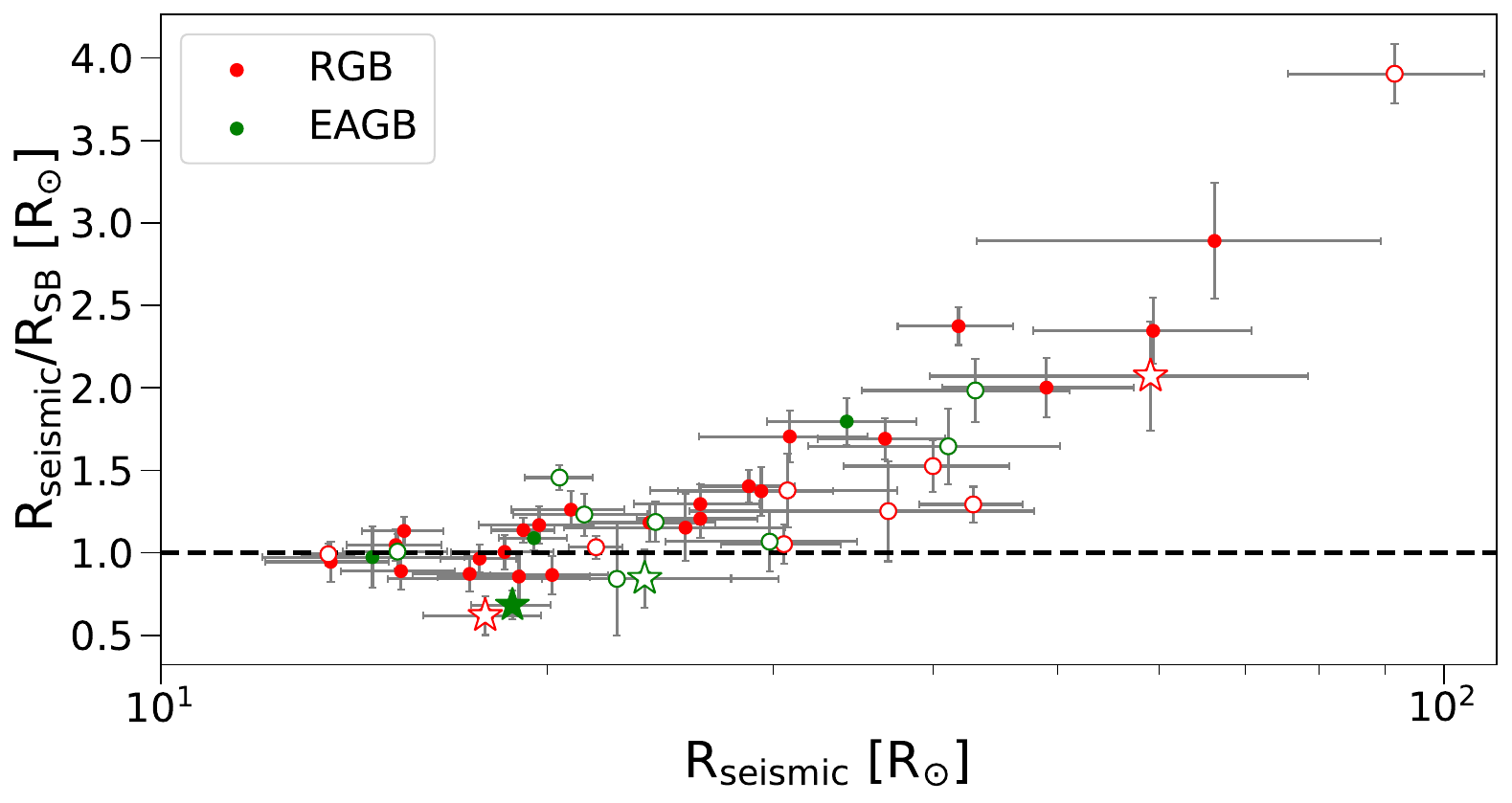}
	\caption{Comparison of the seismic radii and radii calculated from the Stefan-Boltzmann law. Evolutionary status is indicated by the colour coding, and open circles represent stars with a `MD' quality flag. The median residual was $18\%$ with a $1\sigma$ scatter of $\pm62\%$. We note that at $R_{\text{seismic}}\gtrsim30~\rsun$, the ratio between the methods diverges to large values.}
	\label{fig:Radii}
\end{figure}

The seismic radii were compared to an independent radius estimate calculated using the Stefan-Boltzmann (SB) law, $R_{\text{SB}} = \left[L/(4\pi T^4)\right]^{1/2}$. The median residual between the seismic and SB radii estimates was $18\%$ with a $1\sigma$ scatter of $\pm62\%$ (Fig.~\ref{fig:Radii}). This is significantly larger than the radius comparison in the \citetalias{Howell22_M4} study for M4, with a median residual and scatter of $2\%$ with a $1\sigma$ scatter of $\pm20\%$. The large scatter in the M80 radii is mainly driven by the stars with large radii ($\gtrsim30~\rsun$), where a deviating trend is observed. 

We speculate that the larger differences in the radii estimates could be attributed to the poorer measurements of $\Delta\nu$ for M80. This is compounded in the seismic radius relation, where there is a large dependence on $\Delta\nu$. This was also seen for \textit{K2} red giants in \citet{Zinn19_radii_systematics}, and at higher radii ($R_{\text{seismic}} \gtrsim 50~\rsun$) in Zinn et al. (in press). They similarly concluded that the properties determined via the asteroseismic parameters were unreliable at high radii. We discuss possible systematic effects in the seismic masses of stars with $R>30~\rsun$ in Section~\ref{sec:robustness_numax}. 

We also computed a seismic surface gravity using the scaling relation:
\begin{equation}
    g/g_{\odot} \simeq \left(\frac{\nu_{\text{max}}}{\nu_{\text{max},\odot}}\right)\left(\frac{T_{\text{eff}}}{T_{\text{eff},\odot}}\right)^{-1/2}
\end{equation}
For our sample of stars we found an average random uncertainty of $\pm0.03~\mathrm{dex}$ and a systematic uncertainty of $\pm0.004~\mathrm{dex}$. Using \textit{VLT} spectra for a sample of M80 stars, \citet{Caretta15_M80} derived spectroscopic $\log({g})$ uncertainties of $\pm0.04\text{(internal)}\pm0.06\text{(sys)}~\mathrm{dex}$. Thus, the seismic and spectroscopic random/internal uncertainties are comparable, whereas the systematic uncertainty is one order of magnitude less with asteroseismology. Comparing the $\log({g})$ estimates for an overlapping sample of 15 RGB stars with the \citet{Caretta15_M80} study, we calculated an average offset and scatter of $\log({g})_{\text{spec}}-\log({g})_{\text{seis}} = -0.10\pm0.03~\mathrm{dex}$. The asteroseismic estimates are  marginally lower compared to the spectroscopic values. We believe that the offset is possibly due to systematics, e.g. \citet{Caretta15_M80} adopt a RGB mass of $0.85~\msun$ which is too large for this cluster (see Section~\ref{sec:mass_results}). However, the offset and scatter is small, providing confidence to either method of estimating $\log({g})$. We have included the final seismic $\log({g})$ estimates in Table~\ref{tab:final_results}.

\section{Mass Results}
\label{sec:mass_results}
\subsection{Seismic masses and their averages}
\label{sec:masses_averages}
Using our values for $\nu_{\text{max}}$, $\Delta\nu$, $T_{\text{eff}}$ and $L$, we calculated four seismic mass estimates for each star using Equations~\ref{eq:mass_relation}-\ref{eq:mass_relation4}. Corrections were applied to $\Delta\nu$ from \texttt{asfgrid} with the recent extension to low metallicities \citep{Stello22_asfgridv2}. This accounts for deviations from the $\Delta\nu\propto\rho^{1/2}$ scaling relation \citep{Stello09_dnu_numax_relation,White11_dnucorr1,Miglio12_OCstudy,Miglio13_dnucorr3}. Although \texttt{asfgrid} does not have explicit $\Delta\nu$ corrections for EAGB stars, we treated them as RGB stars due to their similar stellar structures. 

Using these individual masses, we calculated an average mass for each evolutionary phase for both uncorrected masses and $\Delta\nu$-corrected masses. We implemented an iterative sigma clipping process to remove any statistical mass outliers from the average mass estimates. This involved excluding stars that were 2$\sigma$ away from the mean and was completed twice. We discuss possible explanations for the mass outliers in Section~\ref{sec:mass_outliers}. We adopted the standard error on the mean as the uncertainty on the average mass estimates.  

\begin{table*}
\centering
\footnotesize
\caption{Average masses, $\overline{M}$, calculated for each evolutionary phase using Eqs~ \ref{eq:mass_relation}-\ref{eq:mass_relation4}. The subscript \textit{corr} indicates the mean mass estimates that used the $\Delta\nu$-corrected masses. The average RGB mass using Eq.~\ref{eq:mass_relation3}, $\overline{M}_{3}$, was derived from the mode of the distribution in Figure~\ref{fig:KDE_scatter} (see Sec.~\ref{sec:masses_averages}), and is denoted with an asterisk. The uncertainty is the standard error on the mean. All units are $\msun$. The average number of stars, $N$, used for each calculation after sigma clipping is provided in the last column.}
    \begin{tabular}{lcccccccc}
      \hline 
      Evol Stage & $\overline{M}_1$ & $\overline{M}_{1,\text{corr}}$ & $\overline{M}_2$ & $\overline{M}_{2,\text{corr}}$ & $\overline{M}_{3}$ &  $\overline{M}_4$ & $\overline{M}_{4,\text{corr}}$ & $N$\\ 
      \hline 
      RGB & $1.46\pm0.19$ & $1.01\pm0.09$ & $0.63\pm0.04$ &	$0.73\pm0.04$ & $0.782\pm0.009$* & $1.06\pm0.06$ & $0.93\pm0.06$ & 31\\
      EAGB & $0.82\pm0.11$ & $0.68\pm0.10$ & $0.50\pm0.04$ & $0.58\pm0.05$ & $0.61\pm0.02$ & $0.80\pm0.08$ & $0.70\pm0.07$ & 11\\
      \hline 
    \end{tabular}
\label{tab:average_masses}
\end{table*}


As seen in Table~\ref{tab:average_masses}, which shows a summary of the average mass results, the Equation~\ref{eq:mass_relation3} average masses were found to have the smallest uncertainties. This equation is independent of the highly uncertain parameter $\Delta\nu$, which we do not believe is reliable for our \textit{K2} dataset (see discussion in the final paragraph of Section~\ref{sec:seismic_quantities}). Additionally, mass estimates from Eq.~\ref{eq:mass_relation3} are the most consistent across many studies using \textit{K2} photometry (e.g. \citealt{Miglio_M4_study}, \citetalias{Tailo22_M4}, \citetalias{Howell22_M4}). \textit{Hence, we will follow the methodology from \citetalias{Howell22_M4} of only using the Eq.~\ref{eq:mass_relation3} masses for any further results in this paper.} The final mass estimates for individual stars are reported in Table~\ref{tab:final_results}.

\begin{figure*}
	\centering
	\includegraphics[width=2\columnwidth]{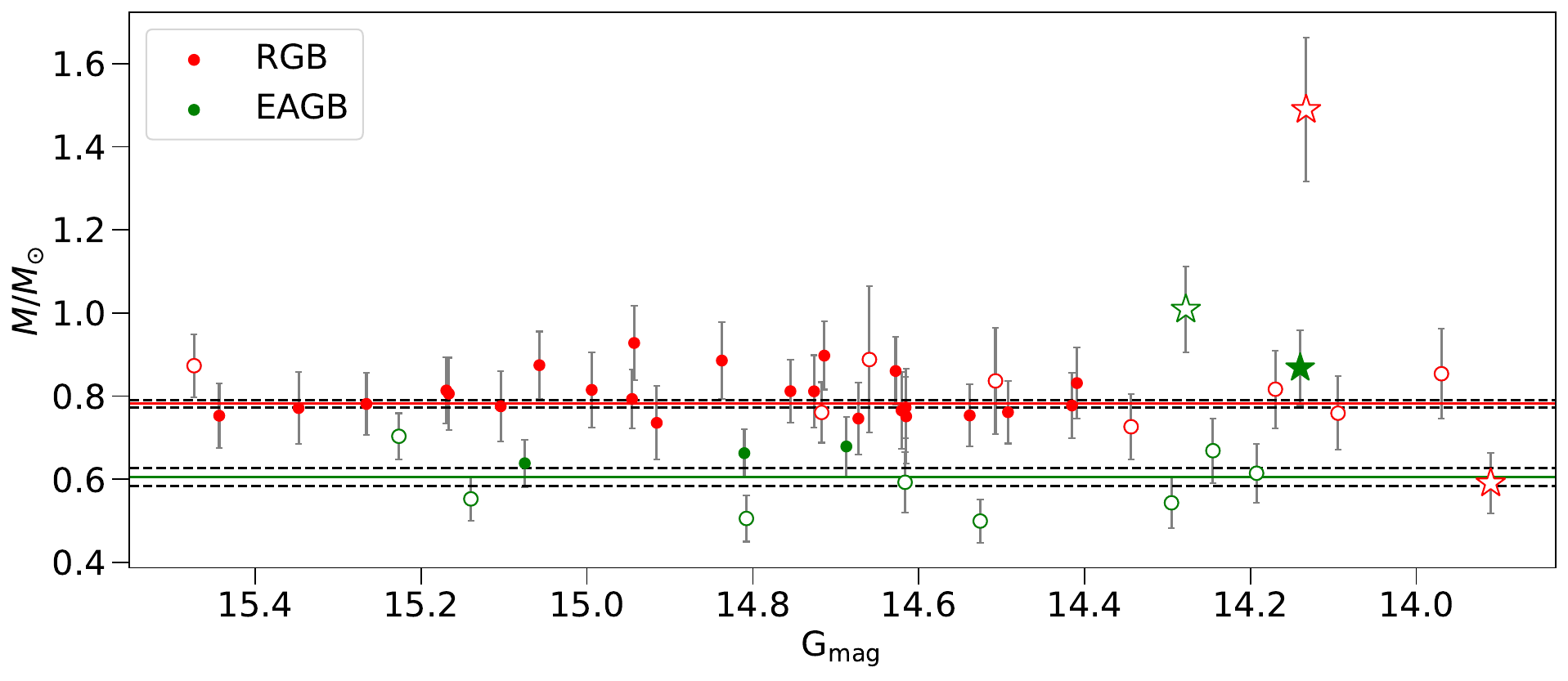}
	\caption{Equation~\ref{eq:mass_relation3} seismic mass estimates for our final sample. Error bars represent the random uncertainties for each individual star (Table~\ref{tab:final_results}). Open symbols indicate stars with marginal detection quality flags (see Section~\ref{sec:seismic_quantities}). The mean EAGB mass (without outliers)  and the mass determined from the mode of the KDE distribution for the RGB (Fig.~\ref{fig:KDE_scatter}) are shown by the solid horizontal lines for each evolutionary phase, and appropriately coloured. The 1$\sigma$ uncertainties (standard error on the mean) of the average masses for each evolutionary phase are indicated by the black dashed lines. Mass outliers (Sec.~\ref{sec:mass_outliers}) are identified by the star symbols.}
	\label{fig:Eq3_mass_distribution}
\end{figure*}

In Figure~\ref{fig:KDE_scatter} we show the mass distributions for the RGB and EAGB samples. We note that the RGB distribution is slightly skewed, and that a straight mean estimate does not represent the most common mass. Thus for the RGB average mass, we use the mode of the distribution ($0.782\pm0.009~\msun$; grey line in Fig.~\ref{fig:KDE_scatter}). Here we have adopted the standard error on the mean for the RGB average as the uncertainty. For the EAGB sample, we observe a bimodal distribution. For our current purpose, we use the mean of the EAGB mass distribution ($0.61\pm0.02~\msun$). We discuss in detail the EAGB bimodality in Section~\ref{sec:multi_pops}. We show the distribution of mass against magnitude in Figure~\ref{fig:Eq3_mass_distribution}, where the average masses are illustrated by the horizontal lines.

\begin{figure}
	\centering
	\includegraphics[width=\columnwidth]{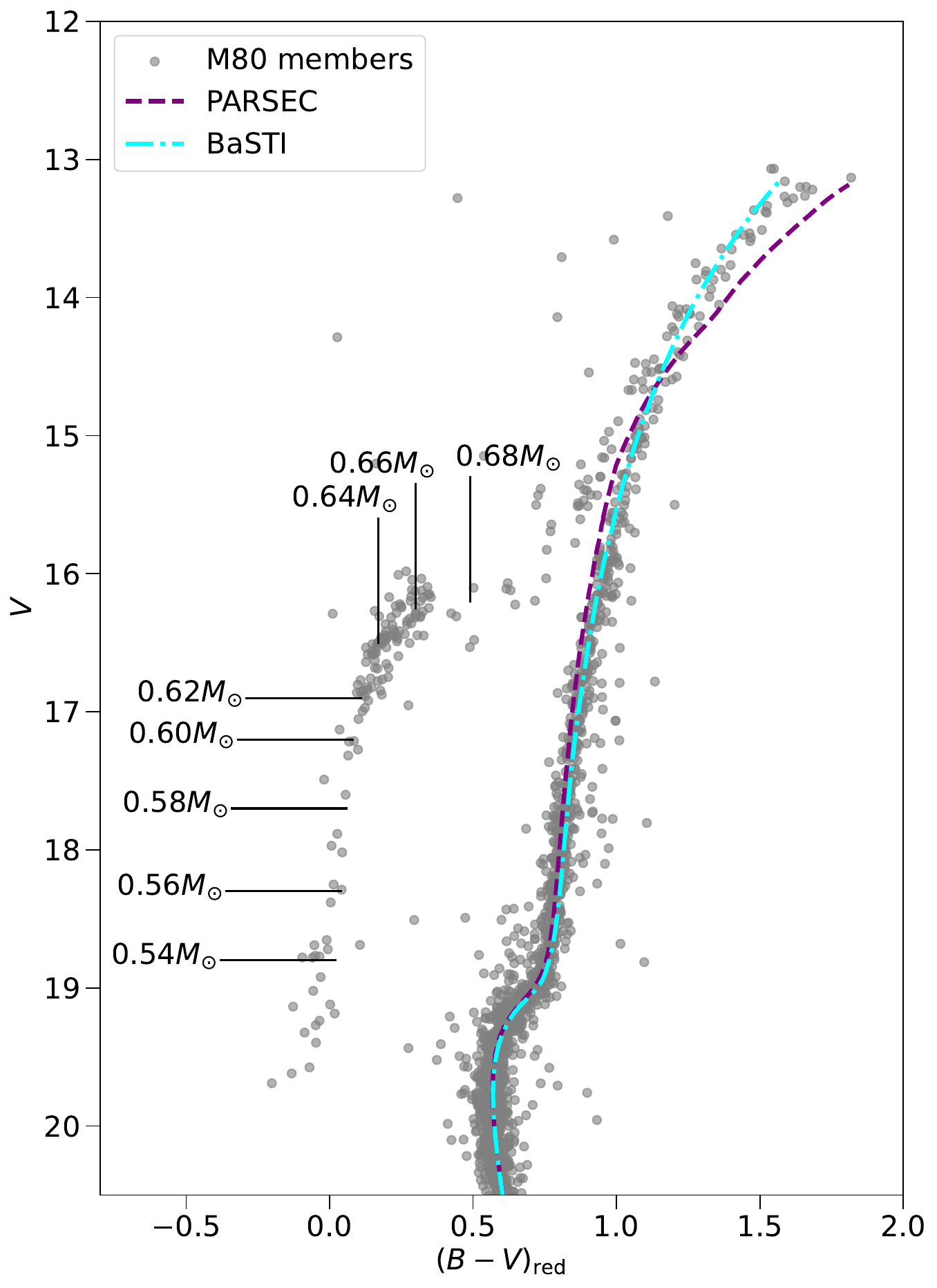}
	\caption{The best fitting  PARSEC and BaSTI isochrones over-plotted on the M80 membership sample from \citet{Vasiliev21_GalacticGC_memberships}. The loci for different HB masses from MESA models (Appendix~\ref{sec:models}) are indicated by black vertical lines. The spread in masses on the HB cannot be described by an initial He abundance variation alone, and requires differing RGB mass loss rates (see Sec.~\ref{sec:second_param_problem}).}
	\label{fig:Isochrone}
\end{figure}

We compared our average RGB mass estimate to RGB initial mass estimates from two different isochrones, PARSEC \citep{Bressan12_PARSEC} and BaSTI \citep{Pietrinferni21_BaSTI}. In Figure~\ref{fig:Isochrone}, we show our best fits to $V,(B-V)$ photometry for each isochrone, with the input parameters summarised in Table~\ref{tab:isochrone_params}.  There is a difference of $0.5$~Gyr between the ages of the isochrones, however each age is within $1\sigma$ of the average reported age for M80: $13.2\pm0.6$~Gyrs\footnote{We derived the average age and scatter for M80 from our literature search \citep{Rosenberg99_M80age1,DeAngeli05_M80age2,Dotter10_M80age3,McDonald15_M80age4,Barker18_M80age5,Valcin20_M80age6}.}. We note that the PARSEC model is not alpha-enhanced, whereas the BaSTI isochrone is ([$\alpha$/Fe]=0.4). To simulate alpha enhancement, we increased the [Fe/H] by $+0.2$~dex \citep{Caretta15_M80} in the PARSEC model. Due to the grid resolution, the BaSTI model also used this metallicity of [Fe/H]~$=-1.6$~dex.

\begin{figure}
	\centering
	\includegraphics[width=\columnwidth]{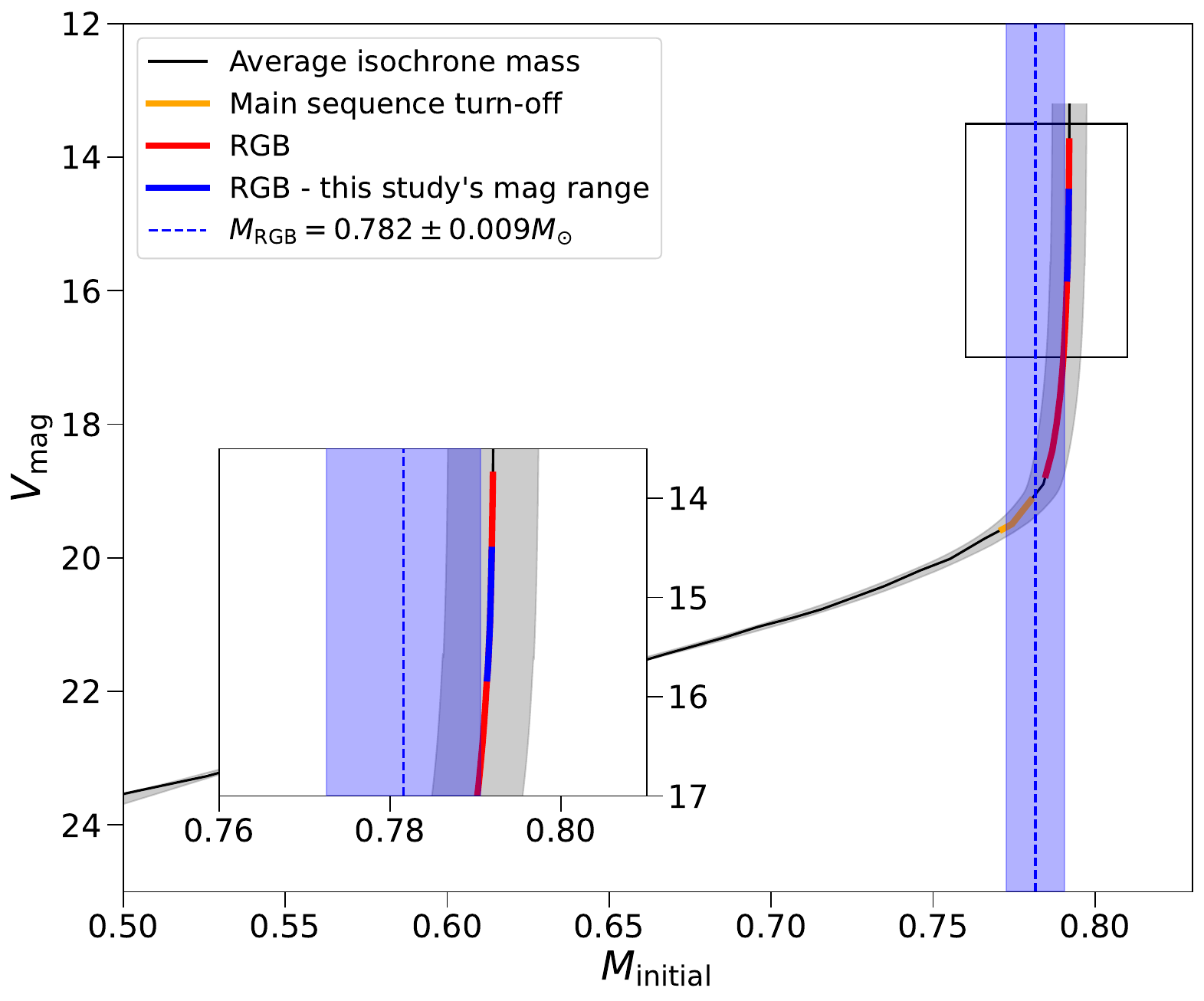}
	\caption{Initial masses against $V$ magnitude (black) derived from the average of the PARSEC and BaSTI isochrones. The adopted uncertainty of the average initial isochrone masses (see Sec.~\ref{sec:masses_averages} for details) is indicated by the grey shaded region. Important evolutionary phases are differentiated by colours; main-sequence turn-off (orange), and RGB (red). The magnitude range of our observational RGB sample are coloured in blue. Our measured average RGB seismic mass is indicated by the vertical dashed line, and the 1$\sigma$ uncertainty is shaded.}
	\label{fig:Isochrone_mass}
\end{figure}

We interpolated between the PARSEC and BaSTI isochrones to determine the average (initial) masses at each point along the evolutionary tracks until the RGB tip. This is shown in Figure~\ref{fig:Isochrone_mass}, which illustrates the evolutionary differences in initial mass. As an indication of the inherent uncertainties in isochrones, we use the original two tracks to define the error range (shaded area in Fig.~\ref{fig:Isochrone_mass}). Comparing the isochrone initial mass for the RGB at a similar magnitude as our sample (blue region), we see that our seismic RGB mass estimate is consistent within the 1$\sigma$ uncertainties. Thus, there is excellent agreement between the models and the asteroseismic masses. This was similarly seen in \citetalias{Howell22_M4} with their results for the lower RGB (e.g. below the luminosity function bump) average seismic mass matching the best-fit M4 model of \cite{MacLean18_chloes_paper}. We also infer that no detectable mass loss has occurred for our RGB stars yet, due to the consistency between the seismic and isochrone initial mass (see further discussion in Sec.~\ref{sec:mass_decline}).


\begin{table}
\centering
\footnotesize
\caption{Input parameters used for the two best fitting isochrone models in Figure~\ref{fig:Isochrone}. The $(m-M)_{\text{red}}$ parameter is the reddened distance modulus, which doesn't account for extinction and is different to the value used to estimate bolometric luminosities in Section~\ref{sec:luminosity}. The initial He content, $Y$, has a small variation between the isochrones, but was the standard value provided in each model. The Reimers' mass loss scaling parameter, $\eta_R$, was kept constant between the isochrones. }.
    \begin{tabular}{lcc}
      \hline 
      Parameter & PARSEC & BaSTI\\
      \hline 
      $[\text{Fe}/\text{H}]$ & -1.6 & -1.6 \\
      $[\alpha/\text{Fe}]$ & 0.0 & +0.4 \\
      Y & 0.2516 & 0.2475 \\
      Age (Gyrs) &  12.5 & 13.0 \\
      $E(B-V)$ &  0.2 & 0.2 \\
      $(m-M)_{\text{red}}$ & 15.75 & 15.75 \\
      $\eta_R$ & 0.3 & 0.3 \\
      \hline 
    \end{tabular}
\label{tab:isochrone_params}
\end{table}

\subsection{Multiple populations: bimodal mass distribution?}
\label{sec:multi_pops}

GCs are known to have variations in their light chemical abundances, revealing the existence of multiple populations. M80 was shown to consist of three chemical sub-populations \citep{Caretta15_M80}, which have different radial distributions throughout the cluster \citep{Dalessandra18_multi_pop_M801}.
Below is our classification and description of the sub-populations of M80, based on these studies:
\begin{enumerate}
    \item Sub-population 1 (SP1): Assumed to be first generation stars with generally scaled-solar compositions. SP1 stars are a factor of $\sim1.5-2$ more numerous within $\sim30''$ of the cluster centre, compared to the other two sub-populations.
    \item Sub-population 2 (SP2): The `intermediate' chemical population which belongs to the second generation of stars, and is enriched in some light metal abundances (e.g. N, Na). SP2 stars are believed to have higher He content, and thus are expected to be lower in mass compared to SP1 (assuming no significant difference in age). A larger fraction of SP2 stars were observed $>30"$ from the cluster centre.
    \item Sub-population 3 (SP3): labelled as the `extreme' second generation stars, with their chemical pattern due to hotter H-burning. Similar to SP2, they are also more numerous $>30"$ of the cluster centre. However, they are the least common overall in the cluster; only 8\% of the \citet{Caretta15_M80} sample consisted of these stars. Their initial masses should be lower than the SP2 stars, because they are inferred to have even higher helium abundances.
\end{enumerate}
From the \citet{Caretta15_M80} spectroscopic catalog, we can determine the expected proportion of each sub-population in our seismic sample. Due to the difference in the radial distributions of the sub-populations, we reduced the \citet{Caretta15_M80} sample to only include stars within a radius of $5'$ of the cluster centre to match the spatial distribution of our \textit{K2} M80 sample. For the 46 stars which had both Na and O abundance measurements in this \citet{Caretta15_M80} sub-sample, $\sim40\%$ belonged to SP1, $\sim60\%$ were SP2 stars, and no stars were found to belong to SP3. Hence, we expect to see a slightly larger proportion of SP2 stars in our seismic sample. If the mass separation between the sub-populations is detectable within our mass uncertainties, then we should observe a bimodal mass distribution. We anticipate that we will not detect a third mass signature corresponding to the SP3, because there are most likely none-to-few stars belonging to this sub-population in our seismic sample.


\subsubsection{RGB sub-populations}
\label{sec:RGB_subpops}
\begin{figure*}
	\centering
	\includegraphics[width=2\columnwidth]{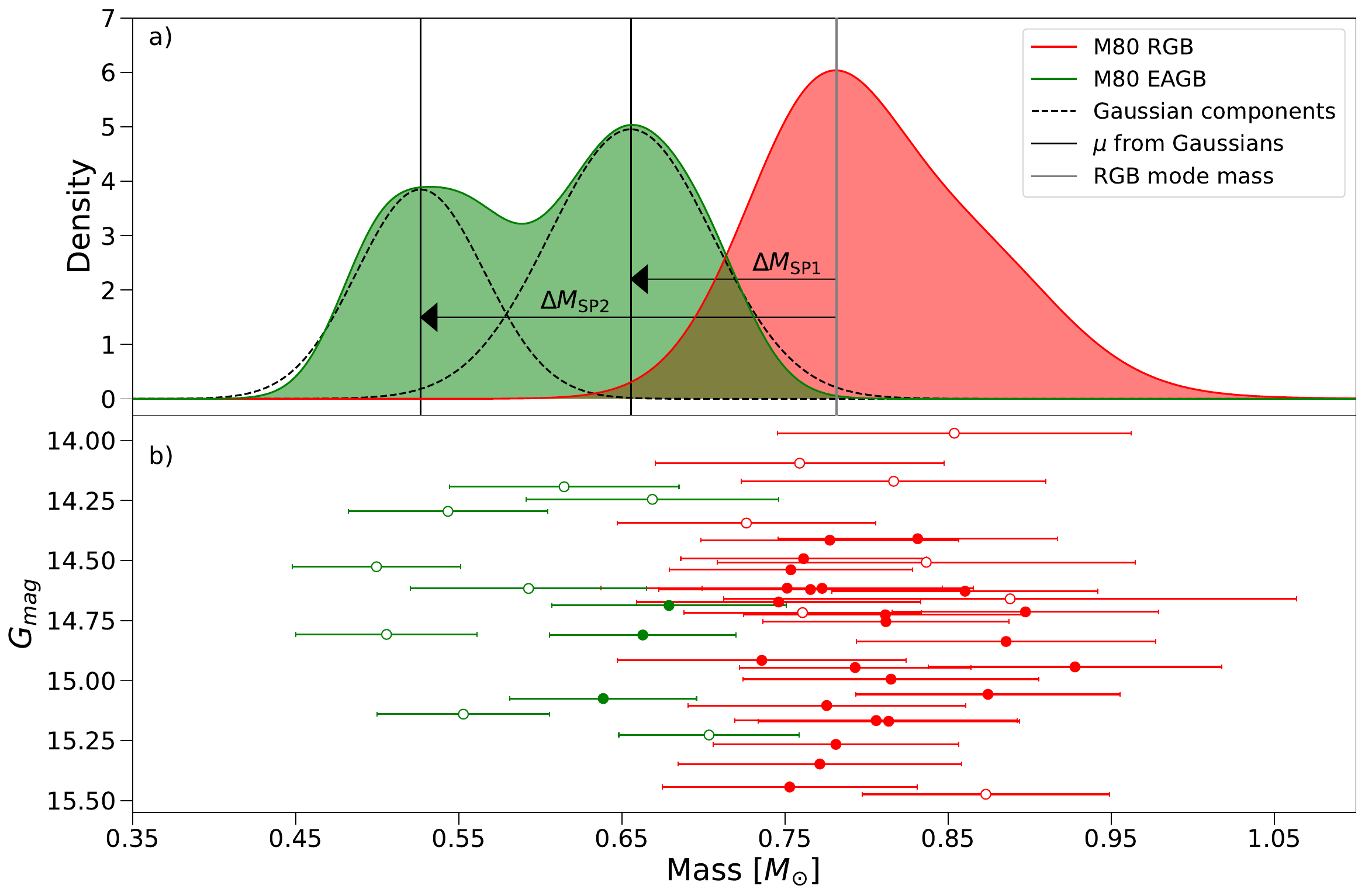}
	\caption{\textbf{(a)} A Gaussian KDE plot showing the mass distributions for the M80 RGB (red) and EAGB (green) samples. The EAGB mass distribution is decomposed into two Gaussian components (dashed black curves), where the means (black lines) represent approximate sub-population masses. The mode of the RGB seismic mass distribution is indicated by a grey vertical line. The arrows illustrate the differences between the integrated mass loss estimates for SP1 and SP2. We note that the mass difference between the sub-populations on the RGB is less than average RGB mass $2\sigma$ uncertainty, hence it is a small systematic. \textbf{(b)} A scatter plot of masses plotted against their \textit{Gaia} $G$ magnitudes. Open circles indicated a MD quality flag. The 1$\sigma$ random uncertainties are included for the masses.}
	\label{fig:KDE_scatter}
\end{figure*}

In Figure~\ref{fig:KDE_scatter} we show Gaussian kernel density estimates (KDE) of our mass distributions. The random mass uncertainties were adopted for the KDE Gaussian widths. In Figure~\ref{fig:KDE_scatter}(b), we have included a scatter plot of the mass estimates against the \textit{Gaia} $G$ magnitudes. Mass outliers were removed from the KDEs (see Sec.~\ref{sec:mass_outliers} and Fig.~\ref{fig:Eq3_mass_distribution}). 

The KDE for the M80 RGB sample has a single peak, which initially suggests that there is only one main mass population. However, this distribution is broad and slightly skewed, with a high-mass tail. To investigate this high-mass tail, we took a representative mass estimate from the tail ($0.86~\msun$), and checked what age this corresponds to using a PARSEC isochrone (see Sec.~\ref{sec:masses_averages} for method). The resultant isochrone had an age of $9.6~$Gyrs, which is inconsistent with the estimated age of $13.22\pm0.58$~Gyrs for this cluster. It also required a reddened distance modulus of $(m-M)_{\text{red}} = 16.0$, which is significantly larger than our estimate of $15.75$. All masses larger than $0.86~\msun$ would have smaller ages and larger distance moduli than the cluster averages. We suggest that the stars giving rise to the high-mass tail of the RGB distribution are instead possible stellar mergers of low mass stars (see discussion in Section~\ref{sec:mass_outliers}).  

Although the RGB mass distribution only has a single peak, there could be a small sub-population mass difference which is being masked by our mass uncertainties. We can use models to estimate the expected mass difference between the RGB sub-populations. To this end, we calculated a small grid of detailed stellar models for M80, which are described in Appendix~\ref{sec:models}. To model SP1, we used an initial He abundance of $Y=0.25$, and adopted an age of $13.0$~Gyrs for the cluster (Sec.~\ref{sec:masses_averages}). We found a modelled SP1 RGB mass of $0.790~\msun$ matched the cluster age. This mass also agrees with our isochrone estimate in Figure~\ref{fig:Isochrone_mass}. The SP1 mass estimate is higher than the peak of our seismic mass distribution for the RGB of $0.782\pm0.009~\msun$ (Fig.~\ref{fig:KDE_scatter} and Sec.~\ref{sec:masses_averages}). A stellar model of this peak RGB mass gives an age of 13.8~Gyrs, too old for M80 (or indeed the Universe). Hence, we hypothesise that our RGB peak mass is the combination of the two (main) populations in M80. To test this, we found a RGB mass estimate for SP2 by increasing the initial He abundance of our model by the average difference in He between the two M80 sub-populations ($\Delta Y = 0.011$; \citealt{milone2018}). We again matched this model to the age of the cluster, and found a RGB mass estimate for SP2 of $0.775~\msun$. Therefore, the mass difference between the sub-populations on the RGB is $0.015~\msun$. Due to the size of our seismic mass uncertainties, we do not expect to detect such a small mass difference in our RGB distribution. Assuming a roughly 50:50 spilt in the sub-populations (Sec.~\ref{sec:multi_pops}), the average mass expected from the models is $0.783~\msun$. This perfectly matches the peak of our RGB mass distribution. We stress that we have not calibrated this at all. Thus, our interpretation is that the main peak of the RGB mass distribution is a combination of these two populations.

In contrast, \citet{Dalessandra18_multi_pop_M801} determined from N-body simulations a significantly larger mass and He abundance difference between sub-populations of $\Delta M = 0.05-0.10~\msun$ and $\Delta Y= 0.05$-$0.06$. This was derived in an attempt to understand the peculiar kinematics observed for M80 cluster members in terms of the properties of sub-populations. If their mass difference is correct, we should detect a signature of a dual mass population in our RGB sample. However, no bimodal distribution was detected. We suggest that a large mass difference or large He abundance variation most likely cannot describe the peculiar kinematics of M80, and alternative theories should be investigated. 

To better quantify the mass difference between the sub-populations in our RGB sample -- and the absolute masses of the sub-populations-- we require spectroscopic data for our entire sample. This would help by i) reducing the individual mass uncertainties with spectroscopic $T_{\text{effs}}$ (see Section.~\ref{sec:teffs_dust}), and ii) classifying our seismic sample into sub-populations with light elemental abundances. Currently, \citet{Caretta15_M80} has the largest spectroscopic abundance catalog of M80 stars (RGB stars only). This sample overlaps by 12 stars with our seismic sample. As a test to see how spectroscopic measurements can help in our analysis, we attempted to use this overlapping sample to see if we can detect a mass difference between the sub-populations. Using the Na abundances, we classified this overlapping sample into SP1 ([Na/Fe]$\lesssim+0.3~\mathrm{dex}$) and SP2 ([Na/Fe]$>+0.3~\mathrm{dex}$)\footnote{The [Na/Fe] value for this split was chosen based on the minimum in the bimodal distribution for the \citet{Caretta15_M80} Na abundances.}. There was a large scatter in the masses (SP1; $\sigma=0.08~\msun$ and SP2; $\sigma=0.06~\msun$), and no significant mass difference was detected between the average masses of the sub-populations ($\Delta M \sim10^{-3}~\msun$). Ideally, we need a larger sample of SP1 and SP2 stars with abundance measurements (preferably our entire seismic sample) to estimate more accurate sub-population masses. This would reduce the uncertainty on the average sub-population masses, due to the larger number of stars. This requires new spectroscopic measurements for this cluster, which we are pursuing.

\subsubsection{EAGB sub-populations}
\label{sec:EAGB_subpops}
For the EAGB masses, we clearly observe a bimodal distribution in Figure~\ref{fig:KDE_scatter}, which suggests that we may be detecting the presence of multiple populations. We decomposed this bimodal mass distribution with two Gaussian fits (dashed curves in Fig.~\ref{fig:KDE_scatter}). From the means of the two Gaussians (black lines in Fig.~\ref{fig:KDE_scatter}), we inferred approximate sub-population masses, assuming that the massive Gaussian corresponds to SP1 and the least massive Gaussian is SP2. The sub-population masses for the EAGB were found to be $0.66\pm0.02~\msun$ (SP1) and $0.53\pm0.02~\msun$ (SP2). It is difficult to quantify the uncertainty of these sub-population masses. Here we adopt the standard error on the mean; we use the 1$\sigma$ standard deviation of each Gaussian, and the number of stars on either side of the minimum in the EAGB distribution (9 stars for SP1 and 4 stars for SP2) as the approximate sample size for each sub-population. 

The difference between the EAGB SP1 and SP2 mass estimates is $0.13~\msun$, which is a factor of $\sim8$ larger than the RGB sub-population mass difference prediction ($0.015~\msun$; Sec.~\ref{sec:RGB_subpops}). This could be evidence of varying stellar mass loss between the multiple populations in GCs. This concept has been investigated by \citet{Tailo19_M4_multipop_massloss} and \citetalias{Tailo20_massloss_difference_multipops}, where SP2 stars are thought to have higher RGB mass loss rates compared to SP1 stars. We cannot determine mass loss \textit{rates} within the limits of our study, but we can measure differences in the integrated mass loss estimates for each sub-population, which could shed light on the second parameter problem. We discuss this in Sections~\ref{sec:SP1_mass_loss} and \ref{sec:second_param_problem}.

\begin{figure}
	\centering
	\includegraphics[width=\columnwidth]{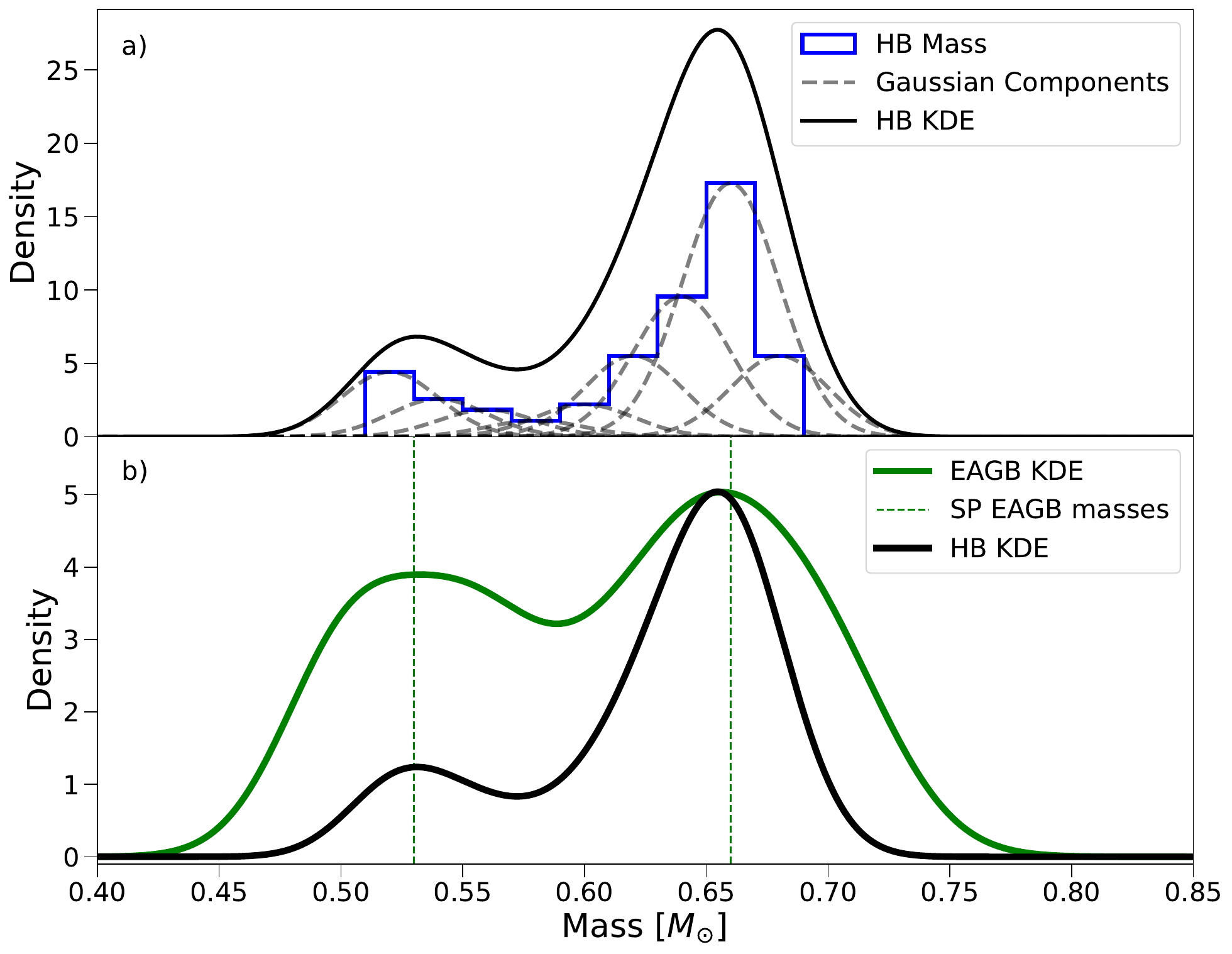}
	\caption{\textbf{(a)} Normalised number density histogram of HB stars in mass bins of width $0.02~\msun$ (blue) inferred from models. Gaussian fits to each bin are shown in grey, where the standard deviation is equal to the bin width. The KDE for the modelled HB mass distribution (black) was produced by summing all Gaussian components. \textbf{(b)} A comparison of the modelled HB KDE (black) and the measured seismic EAGB mass distribution (green; also shown in Fig.~\ref{fig:KDE_scatter}). The modelled HB KDE has been scaled down by a factor of 5.5 for illustrative purposes. The peaks of the modelled HB KDE are similar to the peaks of the EAGB bimodal mass distribution (indicated by green dashed lines, see Fig.~\ref{fig:KDE_scatter}).}
	\label{fig:Modelled_HB_dist}
\end{figure}

Since the EAGB phase directly follows the HB, we would expect the EAGB bimodality to be related to the HB morphology. This can be checked by comparing the mass distributions of each phase. Although we do not have seismic HB mass measurements, we can use models to estimate their masses. We ran a small grid of HB models (Appendix~\ref{sec:models}) and used a \textit{Gaia} HB membership sample (grey points Fig.~\ref{fig:CMD} and \ref{fig:Isochrone}) to build an estimate for the mass distribution of the HB. This HB sample was split into mass bins of size $0.02~\msun$ based on the $V,(B-V)$ locations of the models in the CMD (illustrated in Fig.~\ref{fig:Isochrone}). Figure~\ref{fig:Modelled_HB_dist}(a) shows the resultant normalised number density histograms for each mass bin. We then fit Gaussians to each bin, by using the centre of the mass bin as the mean, and the bin size ($0.02~\msun$) for the standard deviation. A KDE was calculated by summing all the Gaussian components. It can be seen from Figure~\ref{fig:Modelled_HB_dist} that the HB mass distribution was also found to be bimodal. Additionally, the peaks are very similar ($0.65\pm0.03~\msun$ for SP1 and $0.53\pm0.03~\msun$ for SP2) to our sub-population mass estimates from the seismic EAGB mass distribution (Fig.\ref{fig:Modelled_HB_dist}(b)). This is strong evidence that the bimodal distribution that we detected in our EAGB sample is real, given that it matches the inferred mass distribution in the HB. Our EAGB sub-population masses are also consistent within 2$\sigma$ uncertainties to the multiple population HB models for M80 from \citetalias{Tailo20_massloss_difference_multipops}. 

We note that the difference between the heights of the two peaks in the HB mass distribution is larger than that of the seismic EAGB distribution. This could be due to the small EAGB sample size (13 stars). It could also be because HB stars move significantly in the CMD during this phase of evolution (according to models). Lower mass stars are expected to be distributed over a range of magnitudes, which we misclassify as higher-mass stars. This would decrease the amplitude of the lower-mass peak (and increase the amplitude of the higher mass peak). If we were able to directly measure the masses of individual stars in the HB of M80, we could more accurately determine the mass distribution of the HB. 



The quality of the $\nu_{\text{max}}$ measurements could also effect the seismic mass distributions. From Figure~\ref{fig:KDE_scatter}(b), approximately $70\%$ of our EAGB sample have a MD quality flag (illustrated by open circles), which includes all of the least-massive EAGB stars. This could suggest that these stars have inaccurate masses, and as such we could be underestimating the inferred SP2 mass\footnote{We also have a few MD stars in the RGB sample, however it only has a marginal effect on the mass distribution due to the small number of MD quality flags.}. However, the match between the HB and EAGB mass distributions (Fig.~\ref{fig:Modelled_HB_dist}) is evidence that the MD EAGB masses could be correct.

In summary, we present strong evidence that the EAGB mass bimodality in Figure~\ref{fig:KDE_scatter} may be a true detection of the sub-populations. Similar to the RGB results, we need spectroscopic abundances to determine if our bimodal distribution in our EAGB seismic masses is correlated to sub-population membership. \textit{If confirmed, it would be the first direct measurement of a mass difference between sub-populations} (\citetalias{Howell22_M4} only had a tentative detection of a mass difference in their RGB sample).

\subsection{Integrated stellar mass loss for SP1 and SP2 stars}
\label{sec:SP1_mass_loss}
To find an integrated mass loss, we can calculate the difference between the average seismic masses of each evolutionary phase (e.g. \citetalias{Tailo22_M4}, \citetalias{Howell22_M4}). However, we observed a bimodality in the EAGB and HB mass distributions, which most likely corresponds to the multiple populations (Sec.~\ref{sec:EAGB_subpops}). Thus, finding the difference between the average masses reported in Table~\ref{tab:average_masses} would ignore this observed presence of a mass bimodality. Furthermore, as we found with our models (Appendix~\ref{sec:models}) and also in other GC modelling studies (e.g. \citetalias{Tailo20_massloss_difference_multipops}), the sub-populations in M80 need different RGB mass loss rates to reconstruct the observed HB morphology. Therefore, we calculate the integrated mass loss for each sub-population separately.

As mentioned, the magnitude of the HB is too faint to be able to measure asteroseismic signals. Hence, we are unable to determine a purely asteroseismic estimate for the integrated RGB mass loss for M80 with the current \textit{K2} photometry. Instead, we can use our seismic RGB peak mass (see Fig.~\ref{fig:KDE_scatter} and Sec.~\ref{sec:masses_averages}), and the EAGB sub-population masses determined in Section~\ref{sec:EAGB_subpops}, to estimate the integrated mass loss between the RGB and EAGB evolutionary phases, $\Delta M_{RGB-EAGB}$. We note that the mass difference between the sub-populations on the RGB ($0.015~\msun$) is less than average RGB mass $2\sigma$ uncertainty ($0.018~\msun$). Therefore, it is a small systematic in our sub-population integrated mass loss estimates. As illustrated by the arrows in Figure~\ref{fig:KDE_scatter}, our $\Delta M_{RGB-EAGB}$ estimate for SP1 is $0.12\pm0.02~\msun$ and for SP2 is $0.25\pm0.02~\msun$ (see Table~\ref{tab:subpop_massloss} for summary). The uncertainties were derived by the addition in quadrature of the RGB mass and EAGB sub-population mass uncertainties. We discuss the implications of these differing integrated mass losses in Section~\ref{sec:second_param_problem}. 

Although we were unable to measure seismic masses for HB stars, we can use models to estimate the HB sub-population masses. This allows us to estimate an integrated RGB mass loss ($\Delta M_{RGB-HB}$) for each sub-population. Using our HB mass distribution in Figure~\ref{fig:Modelled_HB_dist}, we followed a similar method as described in Section~\ref{sec:EAGB_subpops} of fitting two Gaussians that represent the sub-populations. From the means of these Gaussians, we estimated sub-population masses of $0.65\pm0.03~\msun$ (SP1) and $0.25\pm0.03~\msun$ (SP2). Again, it is difficult to determine the uncertainties on the sub-population massses. Here we adopt the standard deviation of the Gaussian fits. Using these sub-population masses and the seismic RGB mode mass, we find an integrated RGB mass loss, $\Delta M_{RGB-HB}$, for each sub-population of $0.13\pm0.03~\msun$ (SP1) and $0.25\pm0.03~\msun$ (SP2). Mass loss uncertainties were derived using the same method as $\Delta M_{RGB-EAGB}$ estimates. The sub-population masses for the HB and the $\Delta M_{RGB-HB}$ estimates are summarised in Table~\ref{tab:subpop_massloss}.

Our integrated mass loss results for $\Delta M_{RGB-EAGB}$ and $\Delta M_{RGB-HB}$ suggest that there is insignificant mass loss between the HB and EAGB evolutionary phases. This is further discussed in Section~\ref{sec:HB_massloss}. We note that the systematic uncertainties on the average masses have little effect on the integrated mass loss (see discussion in Sec.~\ref{sec:mass_systematics_discussion}).

\begin{table}
\centering
\footnotesize
\caption{Mass estimates for the sub-populations, $M_{\text{SP1}}$ and $M_{\text{SP2}}$, in M80. We adopted the same standard error on the mean for the EAGB sub-population mass estimates, and used the 1$\sigma$ standard deviation of the peaks in the HB distribution (Fig.~\ref{fig:Modelled_HB_dist}, Sec.~\ref{sec:EAGB_subpops}). By taking the differences between these masses from the average RGB mass ($0.782\pm0.009~\msun$), we can calculate integrated mass losses ($\Delta M_{\text{RGB-HB}}$ and $\Delta M_{\text{RGB-EAGB}}$) for each sub-population. The integrated mass loss uncertainties were determined by an addition in quadrature of the individual sub-population uncertainties. Sub-population mass and mass loss estimates that are dependent on models are indicated with an asterisk. All masses in units of $\msun$.}
    \begin{tabular}{lcccc}
      \hline 
       & HB* & EAGB & $\Delta M_{\text{RGB-HB}}*$ & $\Delta M_{\text{RGB-EAGB}}$ \\
      \hline 
       $M_{\text{SP1}}$ &  $0.65\pm0.03$ & $0.66\pm0.02$ & $0.13\pm0.03$ & $0.12\pm0.02$ \\
       $M_{\text{SP2}}$ &  $0.53\pm0.03$ & $0.53\pm0.02$ & $0.25\pm0.03$ & $0.25\pm0.02$ \\
      \hline 
    \end{tabular}
\label{tab:subpop_massloss}
\end{table}



\section{Discussion}
\label{sec:discussion}

\subsection{Mass loss-metallicity trends for SP1}
\label{sec:Mass_loss}
We can directly compare the SP1 integrated mass loss, $\Delta M_{RGB-EAGB}$, for M80 to M4\footnote{The M4 integrated mass losses from \citetalias{Howell22_M4} is for SP1. This is because they are dependent on SP1 mass estimates only, or an average when the expected mass difference between sub-population was small ($\leq0.04~\msun$; \citealt{MacLean18_chloes_paper}).}, by calculating a value from the RGB and EAGB average masses in \citetalias{Howell22_M4}. The $\Delta M_{RGB-EAGB}$ for M4 of $0.28\pm0.01~\msun$ is muchlarger than our M80 mass loss of $0.12\pm0.02~\msun$. For the RGB integrated mass loss, $\Delta M_{RGB-HB}$, we also found that M4 stars lost more mass ($0.17\pm0.01~\msun$ from \citetalias{Howell22_M4} M4 vs $0.13\pm0.03~\msun$ for M80), although there is a smaller difference. We note that the \citetalias{Howell22_M4} M4 study used RGB stars earlier in their evolution in this phase, i.e. stars below the RGB luminosity function bump. We couldn't sample this far down the RGB in M80, because of the faintness of these stars. However, we see from Figure~\ref{fig:Isochrone_mass} that our seismic average RGB mass is consistent with the expected initial mass for RGB stars from isochrones. This suggests that our RGB stars have not experienced significant mass loss yet (this is further discussed in Section~\ref{sec:M80_vs_M4}). The RGB sample used for the mass loss calculation in \citetalias{Howell22_M4} was simiarly identified as having insignificant mass loss. Hence, we are comparing similar RGB samples in terms of mass loss.  

\begin{figure}
	\centering
	\includegraphics[width=\columnwidth]{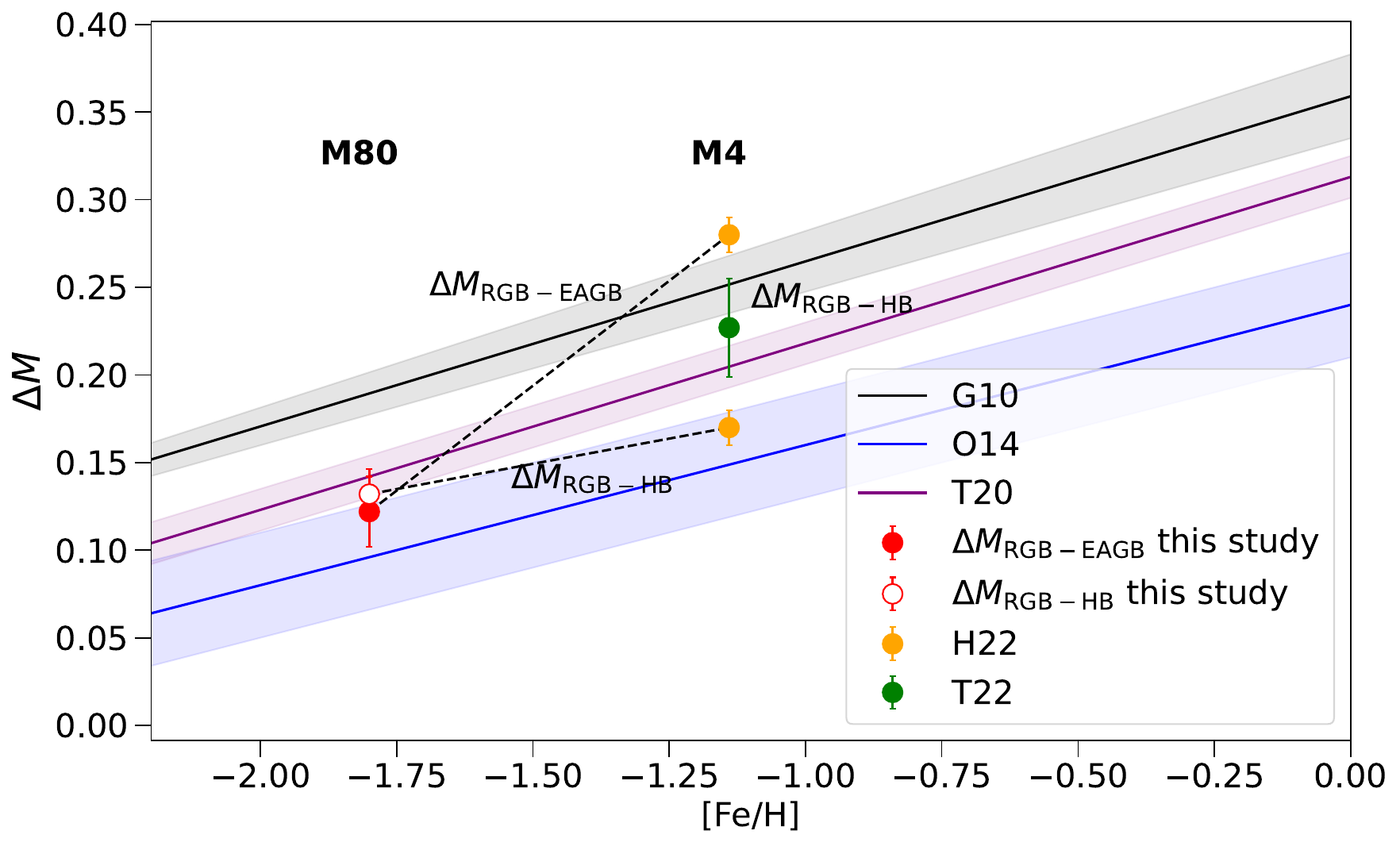}
	\caption{Comparison of the integrated mass loss for SP1 (both $\Delta M_{RGB-HB}$ and $\Delta M_{RGB-EAGB}$) determined from asteroseismic masses for M4 and M80, from this work (red), \citetalias{Howell22_M4} (yellow), and \citetalias{Tailo22_M4} (green). An estimate for the M80 $\Delta M_{RGB-HB}$ partly based on models is included (open circle; see Section~\ref{sec:SP1_mass_loss}). The RGB integrated mass loss ($\Delta M_{\text{RGB-HB}}$) is indicated for the \citetalias{Tailo22_M4} estimate. The integrated mass loss estimates from the current study and \citetalias{Howell22_M4} (M4) are connected by a dashed line, and the $\Delta M_{RGB-EAGB}$ and $\Delta M_{RGB-HB}$ are labelled. This allows us to compare the gradients between the two seismic mass loss estimates to the other mass loss-metallicity trends. RGB mass loss-metallicity trends from \citetalias{Gratton10_masslossGCs} (black), \citetalias{Tailo20_massloss_difference_multipops} (purple), and \citetalias{Origlia14_masslossGCs} (blue) have also been included, with the 1$\sigma$ uncertainty represented by the shaded regions. }
	\label{fig:Mass_loss_comparison}
\end{figure}

It is predicted that metal enriched stars will lose more mass than metal-poor stars. This matches our result where M4 ([Fe/H]~$\sim-1.1$) has a larger stellar mass loss than M80 ([Fe/H]~$\sim-1.8$). We illustrate this in Figure~\ref{fig:Mass_loss_comparison}, where we have plotted different seismic mass loss estimates from this study, \citetalias{Howell22_M4}, and \citetalias{Tailo22_M4}. We have included both the integrated mass loss determined from the mass difference between the RGB and EAGB evolutionary phases ($\Delta M_{RGB-EAGB}$) and the RGB integrated mass loss calculated from the mass difference between the RGB and HB evolutionary phases ($\Delta M_{RGB-HB}$). We also compare these estimates to the RGB mass loss-metallicity trends derived in \citetalias{Gratton10_masslossGCs}, \citetalias{Origlia14_masslossGCs} and \citetalias{Tailo20_massloss_difference_multipops}. 

The M4 $\Delta M_{RGB-HB}$ estimate from \citetalias{Howell22_M4} is consistent within 2$\sigma$ uncertainties with the mass loss-metallicity trend from \citetalias{Origlia14_masslossGCs}, but not the \citetalias{Gratton10_masslossGCs} or \citetalias{Tailo20_massloss_difference_multipops} studies. In contrast, the M4 $\Delta M_{RGB-HB}$ result from \citetalias{Tailo22_M4} agrees within 1$\sigma$ uncertainties to the \citetalias{Gratton10_masslossGCs} and \citetalias{Tailo20_massloss_difference_multipops} mass loss-metallicity trends, but not the \citetalias{Origlia14_masslossGCs} study. The M80 $\Delta M_{RGB-HB}$ (red open circle in Fig.~\ref{fig:Mass_loss_comparison}; partially based on HB models as described in Sec.~\ref{sec:SP1_mass_loss}) and $\Delta M_{RGB-EAGB}$ estimates agree well with the \citetalias{Tailo20_massloss_difference_multipops} mass loss-metallicity trend. Finally, the M4 $\Delta M_{RGB-EAGB}$ from \citetalias{Howell22_M4} is best described by the \citetalias{Gratton10_masslossGCs} relationship (within 1$\sigma$ uncertainty for M80 and 2.5$\sigma$ uncertainties for M4). We note here that all the mass loss-metallicity trends were derived from estimates of the RGB mass loss only, and therefore does not take into account any HB mass loss. Hence, it is a coincidence that the  $\Delta M_{RGB-EAGB}$ from M4 matches the \citetalias{Gratton10_masslossGCs} relation.

There are quantitative differences between the three mass loss-metallicity trends that are included in Figure~\ref{fig:Mass_loss_comparison}. Although the gradients are consistent within $10^{-2}$, there is a $\sim0.1~\msun$ offset between the highest and lowest trends. We think that the difference in the offset is related to the method of measuring the mass loss estimates. 

\citetalias{Gratton10_masslossGCs} and \citetalias{Tailo20_massloss_difference_multipops} both adopted the method of estimating RGB and HB masses from fitting models to photometry, however the application differed between the two analyses. \citetalias{Tailo20_massloss_difference_multipops} considered mass loss differences and He abundance variations between the multiple populations in GCs, and as such their mass loss-metallicity trend was derived from SP1 stars only. Therefore, the \citetalias{Tailo20_massloss_difference_multipops} trend is the most appropriate to compare to. The \citetalias{Tailo20_massloss_difference_multipops} study also reports mass loss differences between the sub-populations, and hence there would be variations in the mass loss-metallicity trend from the same cluster (see Sec.~\ref{sec:second_param_problem} for further discussion). \citetalias{Gratton10_masslossGCs} instead calculated minimum, maximum and median HB masses for each cluster to account for the variation in colour/$T_{\text{eff}}$ on the HBs of GCs. However, for their mass loss-metallicity trend they only used the median HB mass, which does not account for differences in masses and He abundances between sub-populations (although it does match our SP1 HB mass; Section~\ref{sec:SP1_mass_loss}). This appeared to result in an over-estimation of the mass loss, and as they identified, their mass loss estimates are `generally much higher' than others (e.g. from the Reimers' mass loss rates; \citealt{Reimers1975_massloss_rate}). 

\citetalias{Origlia14_masslossGCs} approached measuring the mass loss-metallicity trend differently by correlating the IR colour-excesses of dusty stars with mass loss. They used models of circumstellar dust emission to derive mass loss rates for individual RGB stars. They also included a contribution to the total mass loss from gas, using assumptions of the gas-to-dust ratio and expansion velocities of the stellar envelope. The total RGB mass loss was then determined by integrating the combined dust+gas mass loss rate over the evolutionary time on the RGB. They do state that their mass loss estimates used to derive the mass loss-metallicity trend are lower limits. This is because they adopted a lower limit for the gas-to-dust ratio, and also that mass loss could occur at lower luminosities than what was sampled in their study (which they cannot measure). This may explain why their mass loss-metallicity trend predicts the lowest mass loss in Figure~\ref{fig:Mass_loss_comparison}. We also note that dust production in RGB stars has been suggested to only begin at high luminosities close to the tip ($L > 1000~\mathrm{L}_{\odot}$; \citealt{Boyer10_nodustproduction1,McDonald11_nodustproduction2}). Therefore, mass loss estimates using dust measurements of RGB stars is unlikely to describe the total mass loss during this evolutionary phase. Again, the \citetalias{Origlia14_masslossGCs} mass loss analysis does not consider the multiple populations.

If we can expand our sample of measured integrated mass loss estimates for SP1 from GCs using asteroseismology, we could derive a trend with higher accuracy (as seen from small uncertainties on the seismic masses in Fig.~\ref{fig:Mass_loss_comparison}). This could then be implemented into models to give a more accurate metallicity dependence on the stellar mass loss. This could improve studies which use stellar models as inputs; such as stellar yield calculations in galactic chemical evolution, and population synthesis. However, the metallicity range would only cover the metal-poor regime, and the trend would have to be extrapolated for objects that have metallicities larger than typical GC values\footnote{We note that open clusters could be used as metal enriched seismic targets to study mass loss. However, only a small integrated mass loss was measured with asteroseismology for the open clusters NGC6791, NGC6819, and M67 \citep{Miglio12_OCstudy,Handberg17_OC_study,Stello16_M67}. This is most likely due to the younger ages of open clusters.}. 

\subsection{Further comparisons between M80 and M4}
\label{sec:M80_vs_M4}

\subsubsection{No mass decline on M80 upper RGB}
\label{sec:mass_decline}
The \citetalias{Howell22_M4} study observed a decline in their seismic masses above the RGB bump (refer to their Figure 9) for the GC M4. This suggested that they were seeing the RGB mass loss in action, and that a significant amount of matter is lost directly after the RGB luminosity bump. Initiation of mass loss at the RGB bump was proposed by \citet{Mullan03_bumpmassloss1,Mullan19_bumpmassloss3} who theorised that regions of shear instability at the outer edge of the H-burning shell and the base of the convective envelope during the RGB bump phase could generate small magnetic loops at the surface. This results in mass being lost from the surface in an episodic process. This is unlike the Reimers' or Schr\"oder \& Cuntz mass loss schemes, which scale with luminosity and hence have very low mass loss at the bump magnitude, and are not episodic. \citet{Mullan19_bumpmassloss3} reported that the mass loss rates at the RGB bump could exceed the Schr\"oder \& Cuntz mass loss rates by as much as 3 orders of magnitude. 

Our entire M80 RGB sample is above the luminosity function bump, and as such we are unable to determine if there is a decrease in mass across the RGB bump magnitude. In Figure~\ref{fig:Eq3_mass_distribution}, we also did not observe any decreasing trend in RGB mass vs magnitude. Finally, our measured seismic RGB mass matches the initial mass for upper RGB stars from isochrones (Sec.~\ref{sec:masses_averages}). We suggest that our RGB stars have not lost a detectable amount of mass by this stage. This indicates that the mass loss rate must be very small to this point in the evolution, and that mass loss on the M80 RGB occurs in stars with higher luminosities than our sample. This could be related to the mass loss-metallicity trend where this metal-poor cluster will have a reduced amount of mass loss, even after the bump. To analyse this further, we need seismic masses for RGB stars in M80 at and below the bump, which would require new photometry. 

\subsubsection{HB mass loss \& Mass loss-$T_{\text{eff}}$ dependence}
\label{sec:HB_massloss}
Stars that are currently on the EAGB have recently left the HB phase of evolution. In Figure~\ref{fig:Isochrone}, we showed how mass varies along the HB using a set of models that match the HB morphology (see Sec.~\ref{sec:EAGB_subpops}). We found a range in HB masses of $\sim 0.68$ to $0.52~\rm{M}_{\odot}$. This mass range matched our measured EAGB seismic mass distribution in Figure~\ref{fig:Modelled_HB_dist}. This suggests two concepts: i) we are seeing the masses of two sub-populations being mapped from HB to EAGB, and ii) there is insignificant mass loss on the HB. This agrees with models which use a Reimers' mass loss formulation, because there is very little mass loss at the lower luminosity of the HB. Additionally, \citet{Vink2002_HBmassloss} determined that the mass loss rates for hot (blue) HB stars are too low to show evolutionary effects.

Our findings of insignificant mass loss on the HB (Fig.~\ref{fig:Mass_loss_comparison}) are different to the results found in \citetalias{Howell22_M4} for M4, who measured a seismic HB integrated mass loss of $0.12\pm0.01~\msun$. Their result is in agreement with \citet{Dupree09_RHB_massloss}, who predicts that $0.1-0.2~\msun$ should be lost on the \textit{red} HB. We speculate that the differences in HB mass loss in M80 vs M4 could be due to the metallicity dependence of mass loss (as discussed in Sec.~\ref{sec:Mass_loss}), where metal-poor stars lose less mass than metal-rich stars. 

The effective temperature is also linked to the metallicity of a star. This is especially apparent on the HB, where there is a wide range of $T_{\text{eff}}$. Therefore, effective temperature can be playing a role in HB mass loss. The GC M4 has a different HB morphology compared to M80; it is populated on the red HB and blue HB, whereas M80 has mostly blue HB stars with an extended blue HB tail (see discussion in Section~\ref{sec:second_param_problem}). We suggest that HB stars loose more mass at lower $T_{\text{eff}}$. Thus, the hotter stars on the blue side of the RR Lyrae instability strip will loose less mass than the cooler red HB stars. This could explain why the models and seismic mass estimates infer insignificant mass loss on the hotter HB in M80. In contrast, there was a larger seismically-determined HB mass loss in M4, which has a significant sample of red/cooler HB stars. If true, this implies that a large dependence on $T_{\text{eff}}$ is needed in mass loss formulations used for HB stellar models. 


\subsection{Second parameter problem: is it RGB mass loss?}
\label{sec:second_param_problem}
Metallicity is a key defining factor in the temperature -- and hence location -- of He-burning stars on the HB, however it cannot fully describe the variation in HB morphologies between GCs \citep[see reviews e.g.][]{FusiPecci97_2ndparam_review,Rood97_2ndparam_review2}. Even GCs with similar metallicities and ages are observed to have different HB morphologies. \citet{Milone14_2ndparamproblem} defined two groups of GCs, based on the colour distance between the RGB and red HB, known as the $L_1$ parameter; `M3-like' GCs ($L_1\leq0.35$) have a larger density of stars on the red side of the RR Lyrae instability strip, and `M13-like' clusters ($L_1 > 0.35$) typically have more blue HB stars and exhibit extended blue tails. From this prescription, we can describe M80 as an `M13-like' cluster (see Figures~\ref{fig:CMD}, \ref{fig:UBVI_CMD} and \ref{fig:Isochrone}). The discovery of these different HB morphologies led to the proposition that there is a second parameter besides metallicity -- and most likely a third (e.g.\citetalias{Gratton10_masslossGCs}) -- that determines the temperatures of HB stars. Differing RGB mass loss rates and initial He abundances between sub-populations have been hypothesised as potential second parameters that dictate the HB morphology (\citealt{Rood73_2ndparam,Dorman93_2ndparam2,Dcruz96_2ndparam3,Ferraro98_2ndparam4,Whiteney98_2ndparam5,Catelan00_2ndparam6}; \citetalias{Tailo20_massloss_difference_multipops}). 

In Section~\ref{sec:RGB_subpops} we showed that the expected mass difference between the populations, based on the reported He difference, is small ($\Delta\rm{M}_{SP1-SP2} = 0.015 \rm{M}_{\odot}$). This is consistent with our non-detection of a mass bimodality in the RGB based on our seismic masses (Fig.~\ref{fig:KDE_scatter}). In contrast, we also showed that observationally the EAGB (and HB, via models; Fig.~\ref{fig:Modelled_HB_dist}) has a much larger mass difference between the sub-populations by a factor of 8 ($\Delta\rm{M}_{SP1-SP2} = 0.13 \rm{M}_{\odot}$). Thus, it appears that the HB morphology of M80 cannot be explained by He variation alone. In the context of our models, the HB morphology can only be reconstructed with models that have varying RGB mass loss rates.  \emph{This implies that differing mass-loss rates between the subpopulations must be the main second parameter for M80}.

To quantify the difference in mass loss between the sub-populations, we calculated the seismically-derived integrated mass-loss, $\Delta M_{\text{RGB-EAGB}}$ (Sec.~\ref{sec:SP1_mass_loss}). For each sub-population, we find a $\Delta M_{\text{RGB-EAGB}}$ of: $0.12\pm0.02~\msun$ (SP1), and $0.25\pm0.02~\msun$ (SP2; the independent mass loss of each sub-population is illustrated by the arrows in Figure~\ref{fig:KDE_scatter} and summarised in Table~\ref{tab:subpop_massloss}).

As a comparison, \citetalias{Tailo20_massloss_difference_multipops} calculated integrated mass loss estimates for 46 GCs, including M80, using RGB mass estimates from isochrones and HB mass estimates from model fits to \textit{HST} photometry. Our sub-population integrated mass loss estimates match quantitatively with their M80 modelled values of $\Delta\rm{M} = 0.16\pm0.02~\msun$ (SP1) and $\Delta\rm{M} = 0.27\pm0.03~\msun$ (SP2)\footnote{In Section~\ref{sec:HB_massloss}, we explained how M80 stars do not appear to loose much mass on the HB. Therefore, we can directly compare our $\Delta M_{\text{RGB-EAGB}}$ to the RGB integrated mass loss estimates (mass differences between the RGB and HB) from \citetalias{Tailo20_massloss_difference_multipops}.}. 
\emph{Thus SP2 stars in M80 appear to have a much larger integrated mass loss compared to SP1 stars -- by a factor of about 2}.

\citetalias{Tailo20_massloss_difference_multipops} analysed the effects of the age, metallicity and helium uncertainties on their integrated mass loss estimates. They found that the age uncertainty produced an integrated mass loss uncertainty of $0.013~\msun$, and the metallicity and He abundance uncertainties each resulted in a mass loss uncertainty of the order $10^{-3}~\msun$. These are within their formal mass loss uncertainties of $0.02~\msun$. We also studied the systematic effects of other parameters on our seismic mass estimates (see Section~\ref{sec:mass_systematics_discussion}) and found that there were insignificant effects on the estimated mass differences between evolutionary phases. 



Our work, by using direct seismic mass measurements, adds substantial weight to the theory that mass loss is indeed the (main) second parameter describing HB morphology. If the EAGB mass bimodality is confirmed with spectroscopy (see Sec.~\ref{sec:EAGB_subpops}), we are confident that differing RGB mass loss between sub-populations can resolve the second parameter problem. It may also help in explaining the `missing' SP2 AGB star problem, where SP2 stars are found to be less common on the EAGB than the RGB of some GCs (eg. \citealt{norris81_6752,Campbell2013_NGC6752,MacLean18_chloes_paper,Wang2017_noSP2AGB}). Therefore, we speculate that increased mass loss in SP2 stars could increase the rate of stars avoiding the AGB (AGB-manqu\'e; \citealt{Dorman1995_agb_manque}).

\subsection{Robustness of $\nu_{\text{max}}$ at low metallicities}
\label{sec:robustness_numax}
A systematic error in the seismic masses could be introduced by the use of the seismic scaling relation for our metal-poor stars. The low-metallicity regime has not been well studied in the context of asteroseismology. Large asteroseismic surveys such as the APOKASC catalogs (APOKASC-1; \citealt{Pinsonneault14_APOKASC1}, and APOKASC-2; \citealt{Pinsonneault18_APOKASC2}) contain $<1\%$ of stars with a metallicity [Fe/H]~$\leq-1.0$, and the K2GAP study \citep{Zinn20_K2GAP_DR2} includes $<4\%$ in the same metallicity regime. Furthermore, these large surveys do not contain many luminous evolved giants, i.e. RGB and EAGB stars with low surface gravities such as in our M80 sample. Hence, the seismic scaling relations have not been robustly tested in the low-metallicity regime or for luminous evolved giants. Due to this, \citet{Pinsonneault18_APOKASC2} concluded that caution should be taken when measuring the masses of stars with low surface gravities and metallicities. They reported that their metal-poor stars appeared to be over-estimated in mass. This is because the measured masses were larger than expected for the metallicities and ages, i.e they expected to measure masses $\sim0.8~\msun$ but instead found masses $>1~\msun$. This was also seen in \citet{Epstein14_metal_poor_seismo} and \citet{Casey18_metal_poor_seismo} who measured seismic masses for 3 and 9 (respectively) metal-poor red giants in the \textit{Kepler} field. We note that these studies only calculate one seismic mass estimate per star (using Equation~\ref{eq:mass_relation}), which is dependent on $\Delta\nu$. We instead use seismic mass estimates that are independent of $\Delta\nu$, and find values that are consistent with models/isochrones (see Section~\ref{sec:masses_averages}).

Systematics in the seismic radii estimates were discussed in \citet{Zinn19_radii_systematics} for stars with radii of $R>30~\rsun$. This was also observed in our seismic sample, where there is a diverging trend at higher radii when comparing to a non-seismic radius estimate (Fig.~\ref{fig:Radii}). One hypothesis that we discussed was that this trend could be caused by the measurement of $\nu_{\text{max}}$ and/or $\Delta\nu$, or systematics in the seismic scaling relations for our sample. Looking at $\Delta\nu$-independent Eq.~\ref{eq:mass_relation3} mass estimates (Fig.~\ref{fig:Eq3_mass_distribution}), we cannot detect any diverging trend for the brighter stars (i.e. stars with larger radii). Hence, we conclude that the diverging trend observed in the radius is most likely introduced by $\Delta\nu$.

The $\nu_{\text{max}}$-based RGB seismic masses are also consistent with isochrone mass estimates in the \citetalias{Howell22_M4} study of M4. This implies that there are no systematic issues in the masses when using the $\Delta\nu$-independent Eq.~\ref{eq:mass_relation3} scaling relation. It is also strong evidence that we can accurately measure $\nu_{\text{max}}$ for evolved GC stars with \textit{K2} photometry. Therefore, our work shows that the Equation~\ref{eq:mass_relation3} scaling relation can be used to measure reliable masses for low metallicity (and low surface gravity) stars. This will be of use for future studies of low metallicity stellar populations, for example Galactic halo stars.

\subsection{Possible systematics in the mass estimates introduced by non-seismic parameters}
\label{sec:mass_systematics_discussion}




Mass systematics can also be introduced by other non-seismic input parameters. We investigated the systematic effects of three parameters, which are required to estimate seismic masses: $E(B-V)$, [Fe/H], and $(m-M)_0$. 

The extinction parameter, which we adopted as $E(B-V) = 0.18$, is used to remove reddening in the $V-K$ colour for the $T_{\text{eff}}$ and the $V$ magnitude for the luminosity. As a test, we instead used an extinction of $E(B-V)=0.22$ measured for M80 by \citet{Recio_Blanco06_GC_catalog}. This only changed the average masses by $10^{-3}~\msun$, which is well within the uncertainties of the individual seismic masses. The same order of magnitude deviation was observed when varying [Fe/H] by 2$\sigma$, using the systematic error on the metallicity from \citet{Caretta15_M80}. We next measured the systematic effects introduced by the true distance modulus. We noticed that a $1\sigma$ variation in the distance modulus caused a shift of $\sim0.05~\msun$ to the masses. The distance modulus for M80 is not well constrained (see the scatter in \citealt{Baumgardt21_GCs_membership}), and hence we concluded that this parameter is the main contributor to the mass systematics. However, this parameter introduces a systematic offset to the absolute masses, and the mass differences remain constant. Hence, we are confident in our final integrated mass loss results.  

\subsection{Mass outliers}
\label{sec:mass_outliers}
Asteroseismology has been used to detect and analyse the stellar products of non-standard evolutionary events both in open clusters \citep{Brogaard12_seismo_eBSs_obs1, Corsaro12_seismo_eBSs_obs2, Handberg17_OC_study,Brogaard21_non_standard_evol_seismo} and recently in under-massive He-burning field stars \citep{Li22_nature}. These stars are identifiable by their irregular masses; either more massive that expected indicating merger or mass accretion events, or under-massive suggesting a mass transfer event has occurred. Within our sigma clipping process (see Sec.~\ref{sec:masses_averages}), we removed 4 stars with outlying masses in M80: 3 massive stars and 1 under-massive star. These stars can be identified in Figure~\ref{fig:Eq3_mass_distribution} by the star symbols, and could be the result of the aforementioned non-standard evolutionary events. We note that all outlier stars were identified as members from proper motions and parallax measurements. Membership for these stars can be further validated with radial velocities, however we would require spectroscopic measurements because no estimates are available in the \textit{Gaia} catalogs. 

Blue stragglers are massive stars found in GCs, which are the resultant objects from either a merger of two main sequence stars \citep{Hills1976_Mergers_BSs} or mass-transfer in a binary system \citep{McCrea1964_BSs}. During the core H-burning phase of a blue straggler star, it is located slightly above the main sequence turn-off of a GC's CMD. However, evolved blue straggler stars are almost identical to `normal' GC evolved stars, and thus can't be photometrically separated in CMDs. Evolved blue stragglers will only be distinguishable by their mass. The most massive RGB star in our sample, M80RGB101, has a mass of $M = 1.5\pm0.2(\text{rand})\pm0.2(\text{sys})~\msun$. This mass estimate is within the hypothesised mass range, $1.2-1.6~\msun$, of blue straggler stars \citep{Tian06_eBSS_models1, Sills09_eBSS_models2}, and hence is an evolved blue straggler candidate. The other massive stars, M80AGB115 ($M =1.0\pm0.1(\text{rand})\pm0.1(\text{sys})~\msun$) and M80AGB18 ($M =0.87\pm0.09(\text{rand})\pm0.12(\text{sys})~\msun$), have masses lower than the expected blue straggler mass range, but still could be due to a merger of lower mass stars.  Additionally, as discussed in Section~\ref{sec:multi_pops}, RGB stars with masses larger than $\gtrsim 0.86~\msun$ would be too young to be on the RGB in this cluster, and could also be possible merger candidates. If these massive stars are instead the resultant objects of mass accretion in a binary system, they would need a binary companion. Unfortunately, there are no catalogs of binaries for M80 RGB and AGB stars (there are some binary surveys of the extreme blue HB in M80; e.g. \citealt{Moni06_EBHB_binaries1, Moni09_EBHB_binaries2}), and as such we are unable to confirm if any of the massive stars are possibly the products of mass accretion. Another theory is that the massive EAGB stars could be misclassified RGB stars, which could be caused by uncertainties in the $BVI$ photometry. This could explain the mass for the star M80AGB18, although the star M80AGB115 would still have a slightly larger mass compared to the average RGB mass for M80.  


The star M80RGB09 is an under-massive RGB star with a mass of $M=0.59\pm0.07(\text{rand})\pm0.08(\text{sys})~\msun$. This mass is more consistent with the average EAGB mass for M80, and it is likely that this is a misclassified EAGB star. In the \textit{Gaia} CMD (Fig.~\ref{fig:CMD}), this star would be classified as an EAGB star, and indicates a problem with the individual BVI photometry for this star. Alternatively, this lower mass could be the result of a non-standard evolutionary event, like mass stripping in a binary system. However, again we are unable to confirm this theory due to the lack of binary surveys for M80. 

Finally, we would like to comment on the $\nu_{\text{max}}$ measurement, and the associated quality flag. All mass outliers, except for the star M80AGB18, have a marginal detection quality flag (illustrated by the open symbols in Figure~\ref{fig:Eq3_mass_distribution}). This suggests that the $\nu_{\text{max}}$ measurement could be less accurate due to the poor signal-to-noise of the power spectra, leading to inconsistent masses. However, the $\nu_{\text{max}}$ uncertainty is factored into the individual random uncertainty for each star, and each outlier star is $>2\sigma$ away from the seismic averages. This lends significance to these stars being true mass outliers.  

The \citetalias{Howell22_M4} study of M4 also detected seismic mass outliers, and an extensive analysis was conducted to gauge the effects of dust extinction on the random uncertainties of the masses. They were unable to conclusively determine if reddening was offsetting the masses. This is because there is a circular dependence of extinction in the photometric $T_{\text{effs}}$ and luminosity. In the \citetalias{Howell22_M4} study, they used the differential dust map from \citet{Hendricks12_M4_dustmap}. We ran a test which instead used the M4 differential dust map from \citet{Alonso_Garcia12_dustmaps}. This dust map surveyed a larger proportion of the cluster, and hence extinction estimates were available for more stars in the \citetalias{Howell22_M4} sample. Using this map, the massive stars reported in \citetalias{Howell22_M4} decreased in mass to within the $1\sigma$ uncertainty of the average RGB mass. This shows that small variations in the extinction ($\sim0.05$ mag) can have significant effects on the individual seismic masses. We note that the new extinction values did not change the masses of the M4 under-massive stars. Currently, no public differential dust map exists for M80, and an average extinction was used for the calculation of the seismic masses. If there are significant local variations in the dust in front of this cluster, it could be the cause of the observed mass outliers in this study. We emphasise again the need for dust-independent spectroscopic $T_{\text{effs}}$.


\section{Conclusions \& Future Work}
\label{sec:conclusion}

We present the first asteroseismic analysis of the globular cluster M80. It is only the second GC with measured seismic masses. We calculated four seismic mass estimates for a sample of 47 red giants using Equations~\ref{eq:mass_relation}-\ref{eq:mass_relation4}, which were dependent on the parameters $\nu_{\text{max}}$, $\Delta\nu$, $T_{\text{eff}}$ and $L$. The $\Delta\nu$-independent mass relation, Eq.~\ref{eq:mass_relation3}, was found to be the most reliable. This is due to the difficulty in measuring $\Delta\nu$ with \textit{K2} photometry, especially for these low metallicity and low surface gravity stars, and has been similarly seen in other \textit{K2} asteroseismic studies of clusters (e.g. \citealt{Miglio_M4_study}, \citetalias{Tailo22_M4} and \citetalias{Howell22_M4}). 

A summary of the main results from our seismic mass estimates are as follows:

\begin{enumerate}
    \item We did not detect a mass bimodality in our RGB sample (Fig.~\ref{fig:KDE_scatter}). This is in agreement with the expected small mass difference between sub-populations of $0.015~\msun$ from models (Sec.~\ref{sec:RGB_subpops}). We measured an average RGB mass of $0.782\pm0.009~\msun$, which matched the RGB mass estimate from isochrones. Our interpretion is that this average mass is a mix of the SP1 and SP2 initial masses on the RGB.
    
    \item We detected a distinct bimodal mass distribution in our EAGB sample, which is likely a detection of a mass difference between the multiple populations (Fig.~\ref{fig:KDE_scatter} and Sec.~\ref{sec:EAGB_subpops}). We estimated sub-population masses in our EAGB sample of $0.66\pm0.02~\msun$ (corresponding to SP1), and $0.53\pm0.02~\msun$ (SP2). These sub-population mass estimates match the peaks in our HB mass distribution (inferred from CMDs and models; Fig.~\ref{fig:Modelled_HB_dist}), and the sub-populations masses after RGB mass loss from \citetalias{Tailo20_massloss_difference_multipops}. Hence, we believe that the observed bimodality in the EAGB mass distribution is likely real, and due to the multiple populations. Spectroscopic $T_{\text{effs}}$ and light chemical abundances are required to confirm/deny this, and to reduce the individual seismic mass uncertainties. 

    \item Using the sub-population masses for the EAGB sample and the average RGB mass, we calculated a seismic integrated mass loss, $\Delta M_{\text{RGB-EAGB}}$, for each sub-population. For SP1, we found a $\Delta M_{\text{RGB-EAGB}} = 0.12\pm0.02~\msun$ and for SP2 $M_{\text{RGB-EAGB}} = 0.25\pm0.02~\msun$. Using a combination of models and our seismic masses, we also found an RGB integrated mass loss for each sub-population of $0.13\pm0.03~\msun$ for SP1 and $0.25\pm0.03~\msun$ for SP2. These mass loss estimates are summarised in Table~\ref{tab:subpop_massloss}. A comparison our sub-population mass loss results for $\Delta M_{\text{RGB-HB}}$ and $\Delta M_{\text{RGB-EAGB}}$ suggests that HB stars in M80 do not have any significant mass loss. This was also seen in the close match between the EAGB and HB mass distributions (Fig.~\ref{fig:Modelled_HB_dist}). In contrast, \citetalias{Howell22_M4} detected an HB integrated mass loss of $0.12\pm0.01~\msun$ for the GC M4. We infer that this HB mass loss difference is due to the aforementioned metallicity dependence on mass loss. We also speculate that HB mass loss could be strongly dependent on $T_{\text{eff}}$, where stars with higher effective temperatures loose less mass. 

    \item Comparing the M80 seismic SP1 integrated mass loss estimates to the \citetalias{Howell22_M4} results for M4, we find that M80 red giants loose less mass. This agrees with previous studies that measure a mass loss-metallicity trend from GCs using less direct methods (Fig.~\ref{fig:Mass_loss_comparison}). We need to populate the mass loss-metallicity space further with seismic analyses of stellar mass loss in other GCs, which would lead to a more accurate mass loss-metallicity trend. 
   
    \item The difference between our EAGB sub-population mass estimates of $0.13~\msun$ is much larger (by a factor of 8) than expected from the modelled sub-population mass difference for the RGB ($0.015~\msun$), given the reported He difference. This suggests that the sub-populations have different RGB mass loss rates. \citetalias{Tailo20_massloss_difference_multipops} found that SP2 stars have enhanced RGB mass loss compared to SP1 stars, in quantitative agreement with our sub-population mass loss estimates. This is strong evidence that differing RGB mass loss rates is the main second parameter defining the variation in HB morphologies. We emphasise the potential of using asteroseismic mass measurements in combination with spectroscopic chemical abundances to accurately measure mass loss differences between the sub-populations in GCs with different HB morphologies. 
    
    \item In the \citetalias{Howell22_M4} seismic study of the GC M4, a mass decline was observed above the RGB luminosity bump. We did not detect a decrease in mass in our M80 RGB mass distribution. This suggests that mass loss on the M80 RGB must occur in stars with higher luminosities than our sample, again suggesting a metallicity dependence for mass loss. 
      
    \item We identified four mass outliers in our sample (star symbols in Fig.~\ref{fig:Eq3_mass_distribution}). We discussed whether their inconsistent masses could be the product of non-standard evolution, such as binary mergers or mass accretion events. In particular, we identify that the the star M80RGB101, with a mass of $1.5\pm0.2$(rand)$\pm0.2$(sys), is an evolved blue straggler candidate. However, we unable to confirm if these outlying masses are due to binary interactions. The likely explanation for these mass outliers (except for M80RGB101) was that they were misclassified in their evolutionary status, due to poor BVI photometry. We also considered that RGB stars with masses $\gtrsim 0.86\msun$ could be possible merger candidates as well. 

    \item Our study illustrates that we can measure $\nu_{\text{max}}$ well for GC stars, and that the $\Delta\nu$-independent mass scaling relation (Eq.~\ref{eq:mass_relation3}) is robust for the low metallicity (and low surface gravity) regime. This is shown by the agreement between our RGB average mass estimate and isochrones (Fig.~\ref{fig:Isochrone_mass}). Using the same mass scaling relation, \citetalias{Howell22_M4}'s study of M4 also found agreement between their average seismic mass for the RGB and isochrones. This finding will be useful for furture asteroseismic studies of low metallicity populations, for example Galactic halo stars.
\end{enumerate}

In our future asteroseismic studies, we will attempt to further increase the number of seismically analysed GCs to include two other \textit{K2} GCs: M9 and M19. However, this is the limit of possible GCs that can seismically studied with the currently available photometry. This will not change in the near future. Unfortunately, both the \textit{TESS} \citep{Ricker15_TESS} and \textit{PLATO} missions \citep{Rauer14_PLATO} have pixels sizes (21”/pixel and 15”/pixel respectively) that are too large to optimally study dense GCs. There is exciting potential in the \textit{Nancy Grace Roman} telescope (expected launch in 2027), which will observe 0.281 deg$^2$ fields of the galactic bulge in the near-IR, and has a proposed pixel sizes of $0.11"$ \citep{Akeson19_Roman_telescope}. If there are GCs within their selected fields, this could allow us to increase the sample of seismically analysed red giants in GCs \citep[for a detailed discussion see][]{Molnar_2023}. With the small pixel sizes, we would also be able to probe further into the cores of GCs. Finally, we would like to highlight the science collaboration ‘High-precision AsteroseismologY of DeNse stellar field' \citep[HAYDN;][]{Miglio21_HAYDN}, who have recently had success in their proposal for the ESA's Voyage 2050 mission. They are advocating for a dedicated space mission committed to gathering long-period and high-precision data of stellar clusters, to expand the use of asteroseismology as a tool to study stellar evolution. 



\section*{Acknowledgements}
S.W.C. acknowledges federal funding from the Australian Research Council through a Future Fellowship (FT160100046) and Discovery Projects (DP190102431 \& DP210101299), and thanks Dr. Carolyn Doherty for interesting and helpful discussions. This research was supported by use of the Nectar Research Cloud, a collaborative Australian research platform supported by the National Collaborative Research Infrastructure Strategy (NCRIS). D.S. is supported by the Australian Research Council (DP190100666). 

Parts of this research was supported by the Australian Research Council Centre of Excellence for All Sky Astrophysics in 3 Dimensions (ASTRO 3D), through project number CE170100013. 

This publication makes use of data products from the Two Micron All Sky Survey, which is a joint project of the University of Massachusetts and the Infrared Processing and Analysis Center/California Institute of Technology, funded by the National Aeronautics and Space Administration and the National Science Foundation. This paper includes data collected by the Kepler mission. Funding for the Kepler mission is provided by the NASA Science Mission directorate. This work has made use of data from the European Space Agency (ESA) mission Gaia (https://www.cosmos.esa.int/gaia), processed by the Gaia Data Processing and Analysis Consortium (DPAC, https://www.cosmos.esa.int/web/gaia/dpac/consortium). Funding for DPAC has been provided by national institutions, in particular the institutions participating in the Gaia Multilateral Agreement. This research used the facilities of the Canadian Astronomy Data Centre operated by the National Research Council of Canada with the support of the Canadian Space Agency.

This research made use of Lightkurve, a Python package for Kepler and TESS data analysis \citep{Lightkurve}. This work was also made possible by the following open-source \texttt{PYTHON} software: \texttt{asfgrid} \citep{asfgrid_v2}, \texttt{pySYD} \citep{pySYD_chontos}, \texttt{NumPy} \citep{numpy}, \texttt{pandas} \citep{pandas}, and \texttt{Matplotlib} \citep{matplotlib}.

We wish to acknowledge the people of the Kulin Nations, on whose land Monash University operates, and in which the majority of this research was conducted. We pay our respects to their Elders, past, and present. 
\section*{Data Availability}
The main data in this article is available in Table~\ref{tab:final_results}. The light curves will be shared on reasonable request to the author.



\bibliographystyle{mnras}
\bibliography{refs} 




\appendix

\section{Stellar Models}
To aid in the analysis of M80 we computed our own set of stellar models using the MESA 1D stellar evolution code (\citealt{paxton2011,paxton2019}; version 12778). We used the standard MESA nuclear network (`basic.net') and standard equation of state (see \citealt{paxton2019} for details). The Mixing-Length Theory (MLT) parameter was calibrated to match the RGB in the CMD of M80, giving a value of $\alpha_{MLT} = 2.8$. Convective boundary locations were based on the Schwarzschild criterion, extended with exponential overshoot (\citealt{herwig1997}) during the core helium burning phase ($f_{os} = 0.016$). We adopted a metallicty of Z~$=0.0004$, which is equivalent to [Fe/H]~$= -1.8$ with a $+0.2$~dex alpha-enhancement \citep{Caretta15_M80}.

First we calculated a small grid of models varying in mass, assuming an initial He abundance of $Y=0.25$ (first sub-population, SP1). We then searched in this grid to find a star that matches the cluster age of 13.0~Gyr (Sec.~\ref{sec:masses_averages}). We found a star of mass $0.790~\msun$ matches these constraints, in agreement with the isochrone in Figure~\ref{fig:Isochrone}.

To find a model for the second subpopulation (SP2), we need to know the expected He abundance. \cite{milone2018} reports an average difference in He between the two populations of $\Delta Y = 0.011$. Expanding our grid of models with this increased He ($Y = 0.26$) and again matching the age of the GC, we find a second generation mass of $0.775~\rm{M}_{\odot}$. As expected from the low He abundance, this star has a lower mass than the SP1 star. The mass difference is $0.015~\rm{M}_{\odot}$.

To examine the HB (and EAGB) mass distribution (Sec.~\ref{sec:EAGB_subpops}) we calculated another small grid of models. The colour (or $T_{\text{eff}}$) of HB stars is primarily a function of the envelope mass (eg. \citealt{Dorman93_2ndparam2}). This envelope mass can be affected by the initial mass of the star (eg. through differing He content for the same stellar age), or through mass loss on the RGB. To ascertain what the mass distribution on the HB is, we calculated a series of models in which we removed various amounts of the stellar envelope just before the ZAHB. The results are displayed in Figure~\ref{fig:Isochrone}, where it can be seen that the colour becomes `bluer' (i.e. $T_{\text{eff}}$ increases) as HB mass decreases. This allows us to estimate the median HB mass for a comparison to our seismic RGB mass. The peak of our HB colour distribution is at $(B-V) = -0.02$~mag, which is in perfect agreement with the locus of the median colour in \citetalias{Gratton10_masslossGCs}. As can be seen in their Figure~5, this value is very consistent among low-metallicity GCs. Our $0.65~\rm{M}_{\odot}$ HB model matches this colour. This is in agreement with the median mass of $0.64 \pm 0.01~\rm{M}_{\odot}$ found by \citetalias{Gratton10_masslossGCs} and \cite{McDonald15_M80age4}. It is also in agreement with the value found by \citetalias{Tailo20_massloss_difference_multipops} ($0.635 \pm 0.021~\msun$). \cite{McDonald15_M80age4} performed an error analysis on the median mass (their Appendix C3), finding an uncertainty of $0.01~\msun$, which we also adopt here, hence our median HB mass is $0.65 \pm 0.01~\msun$.

We use the various models described above to examine the predicted mass distributions due to the multiple populations (Sec.~\ref{sec:multi_pops}), and in determining an integrated RGB mass loss (Sec.~\ref{sec:SP1_mass_loss}).
\label{sec:models}

\section{Full data table}
Below we provide a table with our final results for the global seismic quantities, stellar properties, and mass estimates from Eq.~\ref{eq:mass_relation3} for our entire M80 sample. 

\begin{table*}
\centering
\footnotesize
\caption{\small{Both our star identifications and the corresponding Gaia DR2 identifications are provided for every star in our seismic sample. We have also included a quality flag (QF) for each star, where `MD' is a marginal detection and `D' is a detection (see Sec.~\ref{sec:seismic_quantities} for more details). Uncertainties for the luminosity, seismic radius, and seismic mass are separated into random and systematic errors, respectively. The random uncertainties are included for the $\log{(g)}$, and there was a constant systematical uncertainty of $\pm0.004$. The photometric $T_{\text{eff}}$ error is $\pm110\mathrm{K}$ for all stars.} }
\begin{center}
\begin{tabular}{ccccccccccc}
\hline
ID        & Gaia DR2 ID         & $\nu_{\text{max}}$ ($\mathrm{\mu Hz}$) & $\Delta\nu$  ($\mathrm{\mu Hz}$) & QF & $T_{\text{eff}}$ & $L/L_{\odot}$     & $R/R_{\odot}$         & $\log{(g)}$   & $M_{3}/M_{\odot}$      \\
\hline
M80AGB108 & 6050422929271884672 & 3.0$\pm$0.2                            & 0.84$\pm$0.19                    & MD & 4638             & 295$\pm$12$\pm$20 & 22.7$\pm$7.7$\pm$0.2  & 1.47$\pm$0.03 & 0.61$\pm$0.07$\pm$0.04 \\
M80AGB111 & 6050422929279669760 & 3.0$\pm$0.2                            & 0.73$\pm$0.06                    & MD & 4591             & 304$\pm$14$\pm$20 & 29.8$\pm$5.1$\pm$0.3  & 1.48$\pm$0.03 & 0.67$\pm$0.08$\pm$0.04 \\
M80AGB115 & 6050422963639983232 & 4.5$\pm$0.2                            & 1.00$\pm$0.09                    & MD & 4613             & 317$\pm$14$\pm$21 & 23.8$\pm$4.0$\pm$0.2  & 1.65$\pm$0.02 & 1.01$\pm$0.10$\pm$0.07 \\
M80AGB17  & 6050422203424752896 & 7.1$\pm$0.2                            & 1.41$\pm$0.04                    & D  & 4893             & 162$\pm$5$\pm$11  & 19.5$\pm$1.2$\pm$0.2  & 1.83$\pm$0.01 & 0.66$\pm$0.06$\pm$0.04 \\
M80AGB18  & 6050422826199654912 & 4.0$\pm$0.2                            & 1.07$\pm$0.03                    & D  & 4649             & 311$\pm$13$\pm$21 & 18.8$\pm$1.3$\pm$0.2  & 1.60$\pm$0.02 & 0.87$\pm$0.09$\pm$0.06 \\
M80AGB195 & 6050422894920240640 & 8.2$\pm$0.4                            & 1.71$\pm$0.07                    & MD & 4935             & 120$\pm$3$\pm$8   & 15.3$\pm$1.4$\pm$0.1  & 1.90$\pm$0.02 & 0.55$\pm$0.05$\pm$0.04 \\
M80AGB196 & 6050421413149970944 & 9.6$\pm$0.4                            & 1.91$\pm$0.19                    & D  & 5023             & 126$\pm$3$\pm$8   & 14.6$\pm$2.6$\pm$0.1  & 1.96$\pm$0.02 & 0.64$\pm$0.06$\pm$0.04 \\
M80AGB20  & 6050422959338994688 & 6.4$\pm$0.4                            & 1.01$\pm$0.06                    & D  & 4881             & 181$\pm$5$\pm$12  & 34.2$\pm$4.6$\pm$0.3  & 1.79$\pm$0.03 & 0.68$\pm$0.07$\pm$0.05 \\
M80AGB201 & 6050422065985765120 & 12.1$\pm$0.2                           & 1.81$\pm$0.05                    & MD & 5046             & 112$\pm$3$\pm$7   & 20.4$\pm$1.2$\pm$0.2  & 2.06$\pm$0.01 & 0.70$\pm$0.06$\pm$0.05 \\
M80AGB22  & 6050423062418192512 & 4.9$\pm$0.4                            & 1.04$\pm$0.04                    & MD & 4810             & 197$\pm$7$\pm$13  & 24.3$\pm$2.7$\pm$0.2  & 1.68$\pm$0.04 & 0.59$\pm$0.07$\pm$0.04 \\
M80AGB43  & 6050421657968528768 & 3.7$\pm$0.2                            & 0.68$\pm$0.07                    & MD & 4768             & 214$\pm$8$\pm$14  & 43.1$\pm$8.0$\pm$0.4  & 1.56$\pm$0.03 & 0.50$\pm$0.05$\pm$0.03 \\
M80AGB77  & 6050422718820027264 & 3.1$\pm$0.2                            & 0.63$\pm$0.08                    & MD & 4711             & 270$\pm$10$\pm$18 & 41.1$\pm$9.1$\pm$0.4  & 1.48$\pm$0.03 & 0.54$\pm$0.06$\pm$0.04 \\
M80AGB84  & 6050422826192630912 & 5.7$\pm$0.4                            & 1.21$\pm$0.06                    & MD & 4950             & 159$\pm$4$\pm$11  & 21.4$\pm$2.5$\pm$0.2  & 1.74$\pm$0.03 & 0.51$\pm$0.06$\pm$0.03 \\
M80RGB09  & 6050422860560369920 & 2.6$\pm$0.2                            & 0.48$\pm$0.10                    & MD & 4700             & 349$\pm$14$\pm$23 & 59.1$\pm$19.3$\pm$0.6 & 1.40$\pm$0.04 & 0.59$\pm$0.07$\pm$0.04 \\
M80RGB101 & 6050422860560631424 & 6.4$\pm$0.4                            & 1.37$\pm$0.06                    & D  & 4554             & 316$\pm$15$\pm$21 & 17.9$\pm$1.9$\pm$0.2  & 1.80$\pm$0.03 & 1.49$\pm$0.17$\pm$0.10 \\
M80RGB103 & 6050422894920197248 & 7.8$\pm$0.4                            & 1.55$\pm$0.06                    & D  & 4705             & 171$\pm$7$\pm$11  & 17.4$\pm$1.7$\pm$0.2  & 1.88$\pm$0.02 & 0.89$\pm$0.09$\pm$0.06 \\
M80RGB105 & 6050422890619475840 & 6.3$\pm$1.1                            & 1.04$\pm$0.07                    & MD & 4605             & 197$\pm$9$\pm$13  & 30.8$\pm$6.7$\pm$0.3  & 1.80$\pm$0.07 & 0.89$\pm$0.18$\pm$0.06 \\
M80RGB107 & 6050422929279995136 & 2.8$\pm$0.2                            & 0.58$\pm$0.01                    & MD & 4454             & 381$\pm$22$\pm$25 & 43.0$\pm$4.0$\pm$0.4  & 1.45$\pm$0.03 & 0.85$\pm$0.11$\pm$0.06 \\
M80RGB110 & 6050422929272145152 & 6.4$\pm$0.2                            & 1.13$\pm$0.06                    & D  & 4670             & 199$\pm$8$\pm$13  & 26.3$\pm$2.8$\pm$0.3  & 1.80$\pm$0.02 & 0.86$\pm$0.08$\pm$0.06 \\
M80RGB113 & 6050422929272127360 & 9.5$\pm$0.4                            & 1.66$\pm$0.07                    & D  & 4746             & 151$\pm$6$\pm$10  & 18.5$\pm$1.7$\pm$0.2  & 1.97$\pm$0.02 & 0.00$\pm$0.09$\pm$0.06 \\
M80RGB117 & 6050422963631436672 & 5.4$\pm$0.7                            & 0.56$\pm$0.04                    & MD & 4654             & 227$\pm$9$\pm$15  & 91.6$\pm$15.9$\pm$0.9 & 1.72$\pm$0.05 & 0.84$\pm$0.13$\pm$0.06 \\
M80RGB120 & 6050422959338984704 & 8.2$\pm$0.2                            & 1.58$\pm$0.05                    & D  & 4739             & 149$\pm$5$\pm$10  & 17.7$\pm$1.2$\pm$0.2  & 1.91$\pm$0.01 & 0.79$\pm$0.07$\pm$0.05 \\
M80RGB123 & 6050422997998353792 & 9.2$\pm$0.7                            & 1.08$\pm$0.04                    & D  & 4756             & 140$\pm$5$\pm$9   & 41.9$\pm$4.3$\pm$0.4  & 1.95$\pm$0.03 & 0.81$\pm$0.09$\pm$0.05 \\
M80RGB124 & 6050423028058445824 & 6.9$\pm$0.2                            & 1.24$\pm$0.07                    & D  & 4691             & 176$\pm$7$\pm$12  & 24.0$\pm$2.5$\pm$0.2  & 1.83$\pm$0.01 & 0.81$\pm$0.08$\pm$0.05 \\
M80RGB127 & 6050423131137641728 & 4.6$\pm$0.2                            & 0.64$\pm$0.07                    & D  & 4597             & 251$\pm$11$\pm$17 & 59.4$\pm$11.5$\pm$0.6 & 1.66$\pm$0.02 & 0.83$\pm$0.09$\pm$0.06 \\
M80RGB128 & 6050423135437362688 & 4.9$\pm$0.2                            & 1.14$\pm$0.06                    & D  & 4666             & 226$\pm$9$\pm$15  & 20.2$\pm$2.1$\pm$0.2  & 1.69$\pm$0.02 & 0.76$\pm$0.08$\pm$0.05 \\
M80RGB129 & 6050423204157037952 & 5.8$\pm$0.2                            & 0.92$\pm$0.05                    & D  & 4689             & 200$\pm$8$\pm$13  & 36.7$\pm$4.1$\pm$0.4  & 1.75$\pm$0.02 & 0.77$\pm$0.07$\pm$0.05 \\
M80RGB132 & 6050423307236068096 & 4.6$\pm$0.2                            & 0.70$\pm$0.07                    & D  & 4635             & 243$\pm$10$\pm$16 & 49.0$\pm$8.4$\pm$0.5  & 1.66$\pm$0.02 & 0.78$\pm$0.08$\pm$0.05 \\
M80RGB141 & 6050516074228833024 & 7.5$\pm$0.2                            & 1.18$\pm$0.05                    & D  & 4721             & 183$\pm$7$\pm$12  & 28.7$\pm$2.5$\pm$0.3  & 1.87$\pm$0.01 & 0.90$\pm$0.08$\pm$0.06 \\
M80RGB190 & 6050422310803556480 & 13.0$\pm$1.0                           & 2.27$\pm$0.09                    & D  & 4834             & 99$\pm$3$\pm$7    & 13.6$\pm$1.5$\pm$0.1  & 2.10$\pm$0.03 & 0.77$\pm$0.09$\pm$0.05 \\
M80RGB194 & 6050422894919142016 & 10.3$\pm$0.7                           & 1.62$\pm$0.07                    & D  & 4717             & 119$\pm$5$\pm$8   & 20.9$\pm$2.1$\pm$0.2  & 2.00$\pm$0.03 & 0.81$\pm$0.09$\pm$0.05 \\
M80RGB199 & 6050422001566171648 & 14.0$\pm$0.9                           & 2.21$\pm$0.04                    & D  & 4846             & 90$\pm$3$\pm$6    & 15.5$\pm$1.1$\pm$0.2  & 2.13$\pm$0.03 & 0.75$\pm$0.08$\pm$0.05 \\
M80RGB203 & 6050422070286663168 & 9.5$\pm$0.4                            & 1.82$\pm$0.09                    & D  & 4701             & 128$\pm$5$\pm$9   & 15.4$\pm$1.6$\pm$0.1  & 1.97$\pm$0.02 & 0.81$\pm$0.08$\pm$0.05 \\
M80RGB220 & 6050422203424750848 & 9.5$\pm$0.7                            & 1.61$\pm$0.06                    & D  & 4750             & 127$\pm$5$\pm$8   & 19.7$\pm$2.0$\pm$0.2  & 1.97$\pm$0.03 & 0.78$\pm$0.09$\pm$0.05 \\
M80RGB222 & 6050422929280010368 & 16.3$\pm$0.4                           & 2.56$\pm$0.05                    & MD & 4823             & 88$\pm$3$\pm$6    & 13.5$\pm$0.6$\pm$0.1  & 2.20$\pm$0.01 & 0.87$\pm$0.08$\pm$0.06 \\
M80RGB234 & 6050422207724342912 & 10.7$\pm$0.4                           & 1.73$\pm$0.03                    & D  & 4811             & 134$\pm$4$\pm$9   & 19.2$\pm$1.1$\pm$0.2  & 2.02$\pm$0.02 & 0.87$\pm$0.08$\pm$0.06 \\
M80RGB236 & 6050516834440053248 & 12.7$\pm$0.7                           & 2.13$\pm$0.07                    & D  & 4874             & 105$\pm$3$\pm$7   & 15.2$\pm$1.3$\pm$0.1  & 2.09$\pm$0.02 & 0.78$\pm$0.08$\pm$0.05 \\
M80RGB41  & 6050421417450307456 & 6.9$\pm$0.4                            & 1.18$\pm$0.06                    & D  & 4728             & 181$\pm$7$\pm$12  & 26.3$\pm$3.0$\pm$0.3  & 1.83$\pm$0.03 & 0.81$\pm$0.09$\pm$0.05 \\
M80RGB45  & 6050421795407434368 & 5.1$\pm$0.2                            & 0.64$\pm$0.15                    & D  & 4656             & 217$\pm$9$\pm$14  & 66.3$\pm$23.1$\pm$0.6 & 1.70$\pm$0.02 & 0.75$\pm$0.07$\pm$0.05 \\
M80RGB46  & 6050421825467565952 & 5.5$\pm$0.5                            & 1.24$\pm$0.07                    & D  & 4643             & 201$\pm$9$\pm$13  & 19.0$\pm$2.6$\pm$0.2  & 1.73$\pm$0.03 & 0.77$\pm$0.09$\pm$0.05 \\
M80RGB50  & 6050422031626060032 & 5.7$\pm$0.4                            & 1.02$\pm$0.06                    & D  & 4670             & 191$\pm$8$\pm$13  & 29.4$\pm$4.1$\pm$0.3  & 1.75$\pm$0.03 & 0.75$\pm$0.09$\pm$0.05 \\
M80RGB56  & 6050422070286270976 & 3.5$\pm$0.2                            & 0.77$\pm$0.04                    & MD & 4532             & 313$\pm$16$\pm$21 & 30.6$\pm$3.3$\pm$0.3  & 1.54$\pm$0.03 & 0.82$\pm$0.09$\pm$0.05 \\
M80RGB58  & 6050422100345530624 & 3.8$\pm$0.2                            & 0.70$\pm$0.05                    & MD & 4584             & 266$\pm$12$\pm$18 & 40.0$\pm$5.9$\pm$0.4  & 1.57$\pm$0.03 & 0.73$\pm$0.08$\pm$0.05 \\
M80RGB68  & 6050422134705283584 & 3.1$\pm$0.2                            & 0.67$\pm$0.13                    & MD & 4567             & 331$\pm$16$\pm$22 & 36.9$\pm$11.1$\pm$0.4 & 1.49$\pm$0.03 & 0.76$\pm$0.09$\pm$0.05 \\
M80RGB69  & 6050422134705282944 & 7.8$\pm$0.7                            & 1.16$\pm$0.08                    & D  & 4774             & 150$\pm$5$\pm$10  & 30.9$\pm$4.6$\pm$0.3  & 1.88$\pm$0.04 & 0.74$\pm$0.09$\pm$0.05 \\
M80RGB72  & 6050422242084085248 & 5.4$\pm$0.7                            & 1.05$\pm$0.09                    & D  & 4651             & 203$\pm$9$\pm$14  & 25.6$\pm$5.0$\pm$0.2  & 1.72$\pm$0.05 & 0.75$\pm$0.11$\pm$0.05 \\
M80RGB85  & 6050422826199652096 & 6.0$\pm$0.2                            & 1.21$\pm$0.02                    & MD & 4664             & 185$\pm$8$\pm$12  & 21.8$\pm$1.1$\pm$0.2  & 1.77$\pm$0.02 & 0.76$\pm$0.07$\pm$0.05 \\
\hline
\label{tab:final_results}
\end{tabular}
\end{center}
\end{table*}



\bsp	
\label{lastpage}
\end{document}